\documentclass[journal]{IEEEtran}

\usepackage[dvipsnames]{xcolor}
\usepackage{graphicx}
\usepackage{cite}
\usepackage{amsfonts}
\usepackage{amsmath}
\usepackage{pifont}
\usepackage{amssymb}
\usepackage[percent]{overpic}
\usepackage[dvipsnames]{xcolor}
\usepackage[export]{adjustbox}
\usepackage[T1]{fontenc}
\usepackage{hyperref}
\usepackage{tabu,multirow}
\usepackage{color}
\usepackage{comment}
\usepackage{booktabs, makecell, tabularx}

\begin{document}
\title{Texture-enhanced Light Field Super-resolution \\ with Spatio-Angular Decomposition Kernels}
%
%
%

\author{Zexi~Hu,
        Xiaoming~Chen,
        Henry~Wing~Fung~Yeung,
        Yuk~Ying~Chung,~\IEEEmembership{Member,~IEEE}\\
        and~Zhibo~Chen,~\IEEEmembership{Senior Member,~IEEE}
\thanks{This work was supported in part by the National Natural Science Foundation of China (No. 62177001), Beijing Natural Science Foundation (No. 4222003), and Research Foundation for Advanced Talents of Beijing Technology and Business University (No. 19008021181).}
\thanks{Zexi Hu, Henry Wing Fung Yeung and Yuk Ying Chung are with the School of Computer Science, University of Sydney, Australia. (Email: zexi.hu@sydney.edu.au, henrywfyeung@gmail.com, vera.chung@sydney.edu.au)}
\thanks{Xiaoming Chen is with the School of Computer Science and Engineering, Beijing Technology and Business University, China. (Email: xiaoming.chen@btbu.edu.cn)}
\thanks{Zhibo Chen is with CAS Key Laboratory of Technology in Geo-spatial Information Processing and Application System, University of Science and Technology of China, China. (Email: chenzhibo@ustc.edu.cn)}
\thanks{Zexi Hu and Xiaoming Chen are co-first authors.}
\thanks{Corresponding author: Xiaoming Chen.}
}

%
%

\markboth{Journal of \LaTeX\ Class Files,~Vol.~14, No.~8, August~2015}%
{Shell \MakeLowercase{\textit{et al.}}: Bare Demo of IEEEtran.cls for IEEE Journals}
%



\maketitle

\begin{abstract}
Despite the recent progress in light field super-resolution (LFSR) achieved by convolutional neural networks, the correlation information of light field (LF) images has not been sufficiently studied and exploited due to the complexity of 4D LF data. To cope with such high-dimensional LF data, most of the existing LFSR methods resorted to decomposing it into lower dimensions and subsequently performing optimization on the decomposed sub-spaces. However, these methods are inherently limited as they neglected the characteristics of the decomposition operations and only utilized a limited set of LF sub-spaces ending up failing to sufficiently extract spatio-angular features and leading to a performance bottleneck. To overcome these limitations, in this paper, we comprehensively discover the potentials of LF decomposition and propose a novel concept of decomposition kernels. In particular, we systematically unify the decomposition operations of various sub-spaces into a series of such decomposition kernels, which are incorporated into our proposed Decomposition Kernel Network (DKNet) for comprehensive spatio-angular feature extraction. The proposed DKNet is experimentally verified to achieve considerable improvements compared with the state-of-the-art methods. To further improve DKNet in producing more visually pleasing LFSR results, based on the VGG network, we propose a LFVGG loss to guide the Texture-Enhanced DKNet (TE-DKNet) to generate rich authentic textures and enhance LF images' visual quality significantly. We also propose an indirect evaluation metric by taking advantage of LF material recognition to objectively assess the perceptual enhancement brought by the LFVGG loss.
\end{abstract}


%
\IEEEpeerreviewmaketitle

\section{Introduction} \label{section:Introduction}

Compared with regular images captured by monocular cameras, light field (LF) images can supply richer information with light rays from multiple angular directions in one single capture. Such a characteristic has facilitated several vision-based measurement applications, \textit{e.g.} material recognition \cite{wangLFRecognition_ECCV2016, luLFRecognition_2019}, 3D measurement \cite{yucer2016efficient, heberUshapeICCV2017, wangOcclusionawareDepthEstimation2015}, salient object detection under complex scenarios \cite{shengLFSaliency_ICASSP2016, zhangLFNet_TIP2020} and anti-spoof face recognition \cite{raghavendraLFFace_TIP2015, jiLFHOG_ICIP2016}, which has achieved considerable improvement compared with other types of sensors, \textit{e.g.} monocular cameras \cite{songEDRNet_TIM2020, sellahewaFace_TIM2010, fangFace_TIM2015}, stereo vision \cite{linStereoDepth_TIM2021} and structured light \cite{lilienblum3DStructuredLight_TIM2014}.

\begin{figure}[t!]
    \begin{center}
        \includegraphics[width=0.34\textwidth,trim=2pt 1pt 1pt 10pt, clip=true]{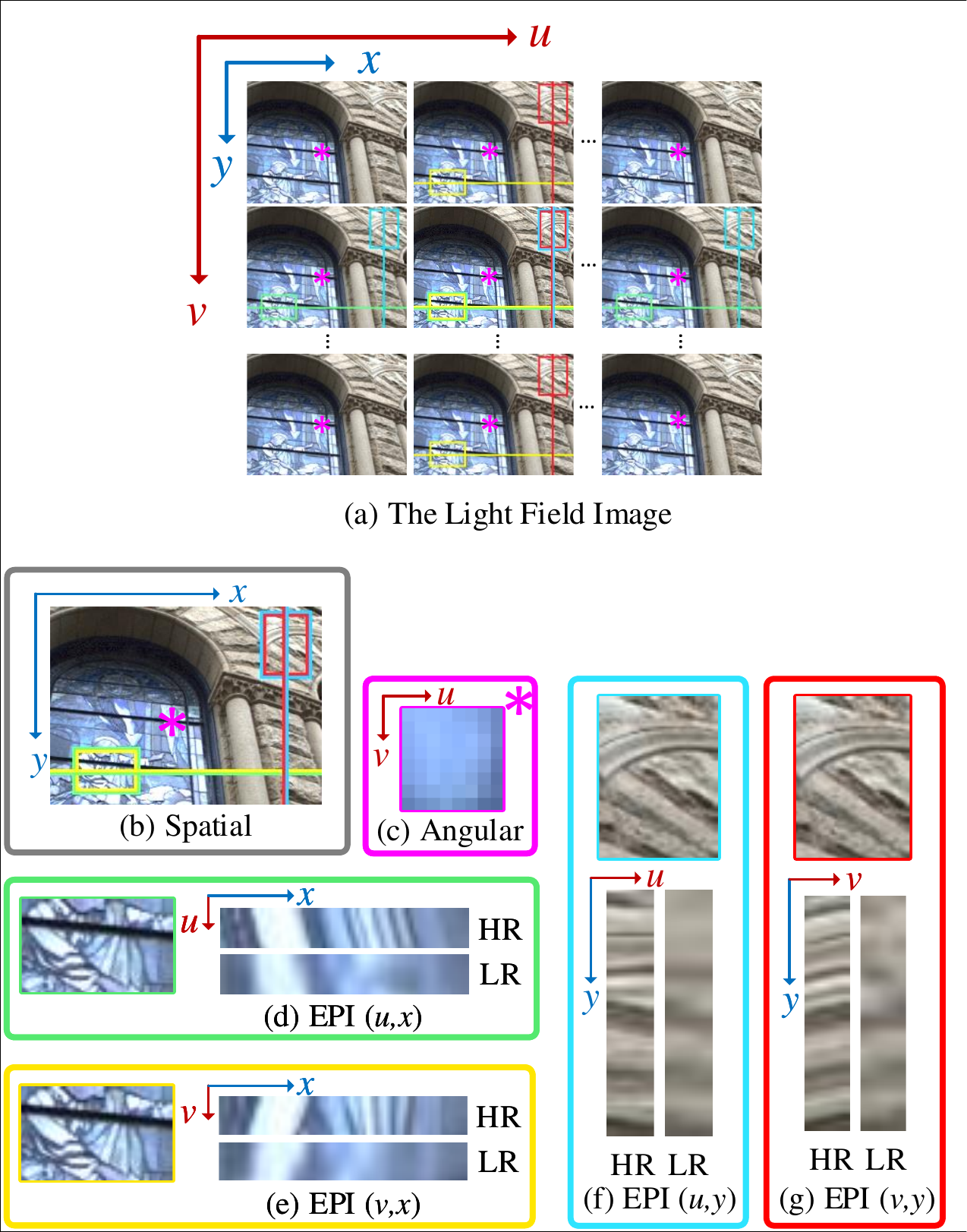}
    \end{center}
    \caption{The sub-spaces of an LF image. LR and HR stand for low and high resolutions. The illustration consists of seven parts: (a) the whole 4D LF image, (b) the spatial sub-space, (c) the angular sub-space, (d)-(g) the EPI sub-space of $(u, x)$, $(v, x)$, $(u, y)$ and $(v, y)$ respectively.}
    \label{fig:Illustration}
\end{figure}

In the past, LF images were usually captured by self-built dense camera arrays \cite{wilburn2004high, wilburn2005high}, which are experimental and expensive for general consumers. With the recent development of more sophisticated LF cameras, \textit{e.g.} Raytrix \cite{Raytrix}, Lytro Illum \cite{Lytro} and Google's Light Field VR Camera \cite{GoogleLF}, LF devices have become increasingly practical in both commercial and industrial usage. However, LF cameras, especially portable LF cameras, usually face a trade-off between the angular and spatial resolutions due to the inherent limitation on the camera's sensor capability \cite{wuLFOverview_JSTSP2017}. Hence the spatial resolution of the images captured from the LF cameras is usually lower than those from the traditional cameras. For instance, a Lytro Illum camera could capture ${14 \times 14}$ sub-aperture images (SAIs) or views, \textit{i.e.} the angular resolution, but each SAI has a low spatial resolution of only ${376 \times 540}$.

To alleviate this problem, a considerable amount of works developed light field super-resolution (LFSR) solutions to increase the spatial resolution of LF images. With convolutional neural networks (CNN), the recent learning-based methods \cite{yoon2017LFCNN, yuanLFSR_SPL2018, zhangResLF_CVPR2019, yeungSAS_LFSR2019} have achieved substantial progress compared with traditional methods \cite{wanner2014variational, rossi2018geometry, ghassab2019light}. The vast majority of these methods are designed to decompose the 4D data structure into sub-spaces of two or three dimensions, which can be optimized with simpler operations, \textit{e.g.} regular 2D and 3D convolutions.

Although these methods have achieved considerable performance with the aforementioned decomposition operations, they are still limited in the following aspects. Firstly, the primary justification of these methods is simplifying the complexity of the 4D data structure and reducing the number of model parameters. The characteristics of the decomposition operations themselves are largely neglected and have not been studied for assisting LFSR. Secondly, due to the sub-optimal architecture design of these methods, their decomposition is confined to limited sub-spaces. Given a LF image shown in Fig. \ref{fig:Illustration} (a), some methods \cite{yeungSAS_LFSR2019, yeungSAS_ECCV2018} can only process the spatial and angular sub-spaces which are shown in the gray and purple boxes of Fig. \ref{fig:Illustration} (b) and (c), and some others \cite{wuEPICNN_TPAMI2018,shinEPINET2018, yuanLFSR_SPL2018, heberUshapeICCV2017} can only process the two typical epipolar-image (EPI) sub-spaces which are shown in green and red boxes of Fig. \ref{fig:Illustration} (d) and (g). Thirdly, the typical form of EPIs reflects insufficient sub-spaces coverage, \textit{i.e.} the sub-spaces shown in the yellow box of Fig. \ref{fig:Illustration} (e) and the blue box of Fig. \ref{fig:Illustration} (f), are long neglected by these EPI-based methods. In fact, these two EPI sub-spaces also carry visual patterns as their siblings do in the green and red boxes, reflecting complementary correlation information from other perspectives. Due to the above limitations, the existing methods cannot extract comprehensive spatio-angular features from their limited decomposed sub-spaces, resulting in sub-optimal performance.

To address this problem, in this paper, we comprehensively discover the potential of the decomposition operations and systematically unify them into the novel concept of "decomposition kernels" to extract spatio-angular features from all the decomposition sub-spaces comprehensively. We propose a novel end-to-end network, namely Decomposition Kernel Network (DKNet), to embed these decomposition kernels. To support a deep network architecture, we employ dense connections and raw image connections to supply long-range information and improve the information flow. The proposed network is experimentally verified to achieve higher PSNR results compared with the previous state-of-the-art methods.

Despite the high performance of DKNet under the PSNR and SSIM metrics, the reconstructed LF images may still be visually unsatisfactory with blurry details caused by the conventional pixel-wise loss functions, \textit{e.g.} the mean square error (MSE). To this end, inspired by the recent perception-based SISR methods \cite{johnsonVGGLoss_ECCV2016}, based on the VGG network \cite{simonyanVGG_2015}, we propose a LFVGG loss to guide the network to focus on perceptual differences. In the experiment, the subjective results show that the Texture-Enhanced DKNet (TE-DKNet) can produce rich textures and visually pleasing SAIs with the guidance of LFVGG loss despite the lower PSNR and SSIM scores. To quantitatively evaluate the perceptual improvement, we also propose an indirect and non-pixel-wise metric by taking advantage of the relevant research on LF material recognition coming with an LF image dataset featuring rich material textures \cite{wangLFRecognition_ECCV2016}. The proposed metric takes both human perception and LF structure into consideration so that it can validate the model's ability to produce authentic textures via the recognition accuracy.

The contribution of this paper is summarized as follows:
\begin{enumerate}
    \item We propose a novel concept of "decomposition kernels" and systematically unify the decomposition operations regarding their sub-spaces into a series of decomposition kernels for comprehensive spatio-angular feature extraction. The end-to-end DKNet incorporates these kernels as well as dense connections and raw image connections for supplying long-range information in achieving superior LFSR performance.
    \item We conduct extensive evaluations to prove that DKNet can achieve state-of-the-art performance surpassing the existing methods. Through comprehensive ablation studies, we delve into the characteristics of proposed components to gain insights into efficient LFSR.
    \item To produce more visual details, we propose a novel LFVGG loss to guide TE-DKNet in texture generation. To quantitatively evaluate the authenticity of generated textures, we also propose an indirect perception-based metric by taking advantage of LF material recognition. Both the subjective and objective results show that our proposed LFVGG loss can guide TE-DKNet to produce realistic textures and enhance visual quality significantly. 
\end{enumerate}

The remainder of the paper is organized as follows. Section \ref{section:Related Works} reviews the related works. Section \ref{section:Methodology} elaborates on the proposed method. Section \ref{section:Experiments} reports the experimental results and analysis. Section \ref{section:Conclusion} presents the conclusion.
\section{Related Works} \label{section:Related Works}
\subsection{Light Field Image Processing}
Conventional methods for LF image processing were usually based on disparity and hand-crafted features. Wanner \textit{et al.} \cite{wanner2014variational} exploited a structure tensor to compute depth maps yielded from disparity maps, and super-resolve the images with the depth map through variational optimization. Wang \textit{et al.} \cite{wang2016simple} developed a simplified variational regularization to reduce the required computation. Rossi \textit{et al.} \cite{rossi2018geometry} proposed to utilize the correlation of LF images by graph-based regularization and obtain the final super-resolved LF images by optimization. Ghassab \textit{et al.} \cite{ghassab2019light} proposed to consider the LFSR problem as two sub-problems, a reconstruction technique and a denoising methodology, and employ the graph-based regularization and the edge-preserving filter to solve these two sub-problems accordingly. Some other methods resorted to coding \cite{perra2016high} and compression \cite{liu2016pseudo, jiang2017light, chen2017light} to tackle the extraordinarily large size of LF data.

With the introduction of deep learning, significant improvement had been achieved for LFSR. Yoon \textit{et al.} \cite{yoon2017LFCNN} proposed to tackle the problem by performing optimization on vertical, horizontal and central pairs of SAI. Similarly, Zhang \textit{et al.} \cite{zhangResLF_CVPR2019} proposed to decompose the problem into multiple sub-networks with different receptive fields of the angular sub-space. Yuan \textit{et al.} proposed an EPI-based method \cite{yuanLFSR_SPL2018} that performed SISR on EPI sub-spaces. EPI-based methods were also proposed for LF reconstruction \cite{wuEPICNN_TPAMI2018} to super-resolve the angular sub-space. However, these methods suffered from poor performance and low efficiency because their network architectures were not end-to-end, which means the sub-networks have no access to other sub-spaces and cannot be optimized jointly. Moreover, their decomposed components need to be operated in a sliding window fashion, leading to duplicated execution costs. 

To address these drawbacks, subsequent works proposed end-to-end neural networks to maximize the merit of supervised learning. Yeung \textit{et al.} \cite{yeungSAS_LFSR2019} proposed a SAS convolution that convolves on the 4D LF data by two 2D convolutions for the spatial and angular sub-spaces respectively. Their experiments demonstrated that SAS convolutions were competent to achieve approximate performance compared to plain 4D convolutions at substantially lower costs of memory and computation. As indicated in \cite{cholletXception_CVPR2017}, separable convolutions can introduce potential non-linearity to the networks, which explained why SAS convolutions can even outperform 4D convolutions on some occasions \cite{yeungSAS_LFSR2019}. Similar success was also achieved in LF reconstruction \cite{yeungSAS_ECCV2018}. However, this work still cannot extract comprehensive spatio-angular features as it has not fully utilized the correlation in all sub-spaces. Based on SAS convolutions, Hu \textit{et al.} further addressed the domain asymmetry issue \cite{huSADenseNet_TIM2021} which commonly exists in LF image processing \cite{wuEPICNN_TPAMI2018}. Meng \textit{et al.} \cite{mengHDDRNet_AAAI2020} used 4D convolution layers to form a high-order residual block to learn geometric features from LF images. Wang \textit{et al.} \cite{wangLfInterNet_ECCV2020} proposed to extract spatial and angular features separately and subsequently perform interaction between them by a series of fusion. To preserve the structure among views, Jin \textit{et al.} \cite{jinLFSSRATO_2020} enforced a structure-aware loss function on their all-to-one neural network.

In general, these works attempted to correlate the spatial and angular information but ended up associating these two domains implicitly and insufficiently. This paper will discover the potential of building explicit connections.

\subsection{Perception Enhanced Super-resolution}
For a long time, PSNR and SSIM have been utilized as the metrics to measure the quality of SISR \cite{kimVDSR_CVPR2016, dongSRCNN_ECCV2014, daiSAN_CVPR2019, lan2020cascading, lan2020madnet, zhou2021cross}. However, some studies \cite{johnsonVGGLoss_ECCV2016, ledigSRGAN_CVPR2017} discovered that these metrics were not necessarily correlated with human perception. Though many recent works have achieved a very high PSNR by training the neural networks with pixel-wise loss functions such as MSE, they still produce blurry results when recovering areas with rich textures. To address this, Johnson \textit{et al.} \cite{johnsonVGGLoss_ECCV2016} discovered that pixel-wise loss functions cannot capture perceptual differences. Instead, they proposed a feature loss that measured the error in the feature space of the VGG network \cite{simonyanVGG_2015}, which was experimentally verified to generate more visual details. Based on this, Ledig \textit{et al.} \cite{ledigSRGAN_CVPR2017} proposed an adversarial loss to discriminate the generated images from the ground-truth and guide the network to produce photo-realistic details. Sajjadi \textit{et al.} \cite{SajjadiEnhanceNet_ICCV2017} extended the loss function with a texture matching loss to learn the correlation of the features, and the experimental results showed that the enhanced details could boost the performance of object recognition which was reasonably associated with human perception. These methods usually achieved lower PSNR but produced images that were more visually appealing. Such a phenomenon has been studied by Vasu \textit{et al.} \cite{vasuEPSR_ECCV2018} as a trade-off between perception and distortion. However, despite the considerable progress of SISR, texture enhancement in LFSR is still undiscovered.

\section{Proposed Method} \label{section:Methodology}
In this section, we will begin with a formal elaboration of the light field sub-spaces in Section \ref{section:LFSubspaces}. Then. we introduce the proposed decomposition kernels in Section \ref{section:DK} and the design of DKNet in Section \ref{section:Network Architecture}. Lastly, we describe the proposed LFVGG loss for further texture enhancement in Section \ref{section:Loss Functions}.

\subsection{Light Field Sub-spaces} \label{section:LFSubspaces}
Formally, the objective of LFSR is to recover a high-resolution (HR) LF image $Y$ from a low-resolution (LR) LF image $X$, denoted as
\begin{equation}\label{eq:obj}
\begin{split}
    \hat{Y} = \mathcal{F}(X) & \quad X(u,v,x,y) \in \mathbb{R}^{U \times V \times W \times H}, \\
    & \quad \hat{Y}(u,v,x,y) \in \mathbb{R}^{U \times V \times rW \times rH}
\end{split}
\end{equation}
where $\hat{Y}$ is an approximation to $Y$, $(U, V)$ and $(W, H)$ denote the LR image's angular and spatial resolutions respectively, and $r$ is the scale. $(u, v)$ denotes an angular location, and $(x, y)$ denotes a spatial location. Generally, 4D LF data can be decomposed into spatial, angular, and EPI sub-spaces, which are visualized in Fig. \ref{fig:Illustration} and briefly described as follows.

\subsubsection{Spatial and angular sub-spaces} Give a 4D LF image shown in Fig. \ref{fig:Illustration} (a), the spatial sub-space, as depicted in the gray box of Fig. \ref{fig:Illustration} (b), is comprised of the two spatial dimensions $(x, y)$ carrying regular 2D image information in individual SAIs. The angular sub-space, marked by the purple box of Fig. \ref{fig:Illustration} (c), on the other hand, is formed by the two angular dimensions $(u, v)$, which encodes the angular information correlating the SAIs. The decomposition into spatial and angular sub-spaces can transform a high-complexity 4D operation into low-complexity 2D operations. For example, Yeung \textit{et al.} \cite{yeungSAS_LFSR2019, yeungSAS_ECCV2018} proposed a spatio-angular separated convolution (SAS) to approximate the 4D convolution by decomposing it into 2D spatial and angular convolutions.

\subsubsection{EPI sub-spaces} On the other hand, an EPI space is comprised of an angular and a spatial dimension to encode the correlation information. Typically, the existing EPI-based methods \cite{wuEPICNN_TPAMI2018,shinEPINET2018, yuanLFSR_SPL2018, heberUshapeICCV2017} decompose the 4D LF data into the sub-spaces $(u, x)$ and $(v, y)$, as depicted in the green box of Fig. \ref{fig:Illustration} (d) and the red box of Fig. \ref{fig:Illustration} (g), and perform optimization on them. In EPI sub-spaces, visual patterns can be observed such as straight lines and sharp edges, and their orientation and distribution also reflect the underlying correlation, \textit{e.g.} the pixel shifts between adjacent SAIs.

Despite the progress achieved by the EPI-based methods, we find that they have neglected the other two EPI sub-spaces, namely $(v, x)$ in the yellow box of Fig. \ref{fig:Illustration} (e) and $(u, y)$ in the blue box of Fig. \ref{fig:Illustration} (f). As shown in the figure, there are visual patterns in these EPIs as well reflecting the correlation from another perspective, and we believe this long-neglected information is as valuable for extracting spatio-angular features as their sibling sub-spaces $(u, x)$ and $(v, y)$. Their impact will be verified in Section \ref{section:Ablation}.

\begin{figure}[t!]
    \centering
    \tabcolsep=0.03cm
    \begin{tabular}{ccccc}
        \includegraphics[width=0.090\textwidth]{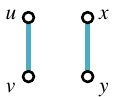}     \vspace{-0.01cm} & 
        \includegraphics[width=0.090\textwidth]{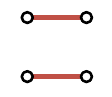}    \vspace{-0.01cm} & 
        \includegraphics[width=0.090\textwidth]{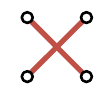}    \vspace{-0.01cm} & 
        \includegraphics[width=0.090\textwidth]{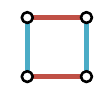}   \vspace{-0.01cm} & 
        \includegraphics[width=0.090\textwidth]{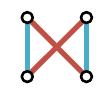}    \vspace{-0.01cm} \\
        \vspace{0.2cm}(a) $\mathcal{K}^{SAS}$ & (b) $\mathcal{K}^{EPI1}$ & (c) $\mathcal{K}^{EPI2}$ & (d) $\mathcal{K}^{\alpha}$ & (e) $\mathcal{K}^{\beta}$ \\
        &
        \includegraphics[width=0.090\textwidth]{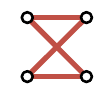}    \vspace{-0.01cm} & 
        \includegraphics[width=0.090\textwidth]{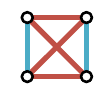}   \vspace{-0.01cm} & 
        \includegraphics[width=0.090\textwidth]{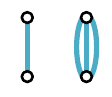}    \vspace{-0.01cm} & 
        \includegraphics[width=0.090\textwidth]{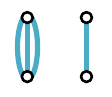}    \vspace{-0.01cm} \\
        &
        (f) $\mathcal{K}^{EPI3}$ & (g) $\mathcal{K}^{\gamma}$ & (h) $\mathcal{K}^{dup1}$ & (i) $\mathcal{K}^{dup2}$\\
    \end{tabular}
    \caption{The decomposition kernels. The red lines represent inter-domain connections, and the blue lines represent intra-domain connections. For simplicity, the dimension notations are omitted except $\mathcal{K}^{SAS}$.}
    \label{fig:Kernels}
\end{figure}

\subsection{Decomposition Kernel} \label{section:DK}
Despite the fact that the implementation of decomposition operations varies in the aforementioned methods, we can unify these operations into decomposition kernels by abstracting the sub-spaces which they are performing on. The unified decomposition kernels can be categorized into the basic SAS-based and EPI-based kernels. Based on these two kernels, we assemble them into two extra categories, namely the combination kernels and the duplication kernels.

\subsubsection{SAS-based Kernel}
A SAS-based kernel decomposed a 4D convolution into a spatial convolution and an angular convolution. It is implemented by firstly reshaping the 4D tensor $X$ into $X^{x,y}\in R^{UV \times W \times H}$, and then employing a regular 2D convolution operator to convolve on the spatial sub-space $(x, y)$. A similar operation is subsequently operated on the angular sub-space $(u, v)$ by reshaping the intermediate tensor into $X^{u,v}\in R^{WH \times U \times V}$ and employing another 2D convolution operator to convolve on $(u, v)$. We formally define the above process as
\begin{equation}
    \mathcal{K}^{SAS}(X)=F^{u,v}(F^{x,y}(X))
\end{equation}
where $F^{d_1, d_2}$ denotes the combined function of reshaping and convolution on the dimensions $(d_1, d_2)$:
\begin{equation}
    F^{d_1, d_2}(X) = f(W \otimes X^{d_1,d_2})
\end{equation}
where $\otimes$ and $W$ are the 2D convolution operation and its corresponding weights, $X^{d_1,d_2}\in R^{d_3 d_4 \times d_1 \times d_2}$ is the reshaped tensor of $X$, and $f(\cdot)$ is the activation function, which is rectified linear units (ReLU) \cite{nairReLU_ICML2010} in this paper.

An illustration of $\mathcal{K}^{SAS}$ is depicted in Fig. \ref{fig:Kernels} (a). As the convolved sub-spaces represent the two intrinsic domains \cite{kalantari2016learning,yeungSAS_ECCV2018,wangOcclusionawareDepthEstimation2015}, \textit{i.e.} the spatial and angular domains, we can say that a SAS-based kernel can builds up two intra-domain connections in a LF image.

However, besides intra-domain connections, we find that there is a lack of inter-domain connections to capture the correlation information across the two domains. In the previous method \cite{yeungSAS_LFSR2019, yeungSAS_ECCV2018}, this drawback can be mitigated to some extent as the inter-domain connections can be implicitly implemented through stacking SAS convolutions. In other words, the connections of the later SAS convolutions can exploit the connections of the preceding SAS convolutions to cross the domains. Nevertheless, we can use EPI-based kernels as a straightforward solution to explicitly build up inter-domain connections.

\begin{figure*}[t!]
    \begin{center}
    \includegraphics[width=0.85\textwidth]{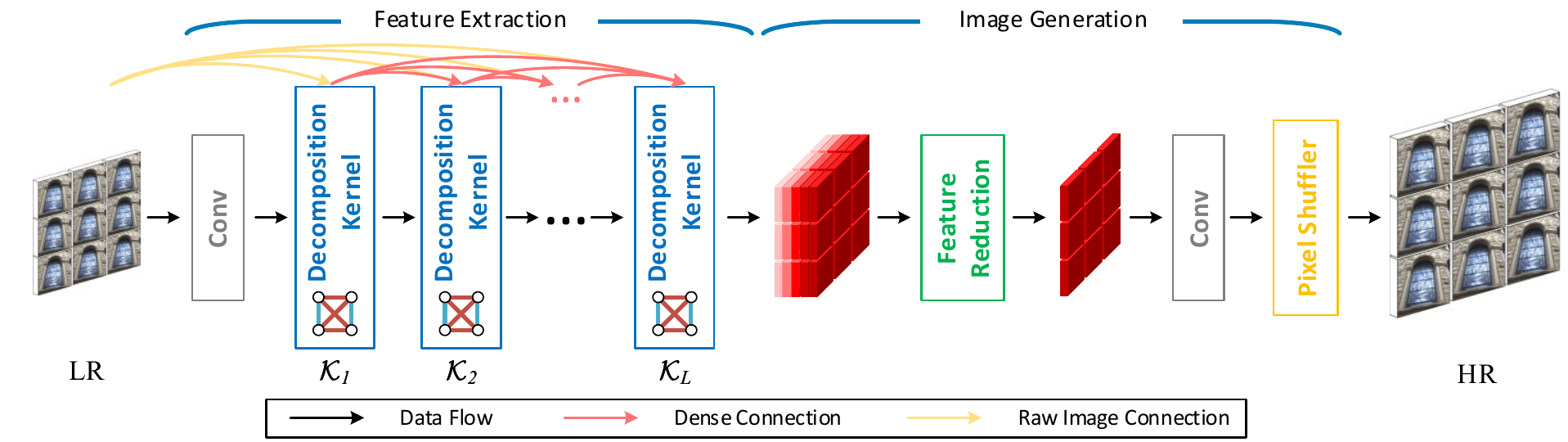}
    \end{center}
    \caption{The architecture of DKNet. The angular resolution of tensors is set to ${3 \times 3}$ for simplicity.}
    \label{fig:Network}
\end{figure*}

\subsubsection{EPI-based Kernels}
Following the implementation of $\mathcal{K}^{SAS}$, we define the first EPI-based kernel $\mathcal{K}^{EPI1}$ to perform on the usual EPI sub-space $(u, x)$ and $(v, y)$ formulated as
\begin{equation}
    \mathcal{K}^{EPI1}(X)=F^{v,y}(F^{u,x}(X)).
\end{equation}

In this way, the optimization on $(u,x)$ and $(v,y)$ can be performed jointly instead of being isolated from each other as in the previous EPI-based methods \cite{wuEPICNN_TPAMI2018,shinEPINET2018, yuanLFSR_SPL2018, heberUshapeICCV2017}. 

Likewise, we define the second EPI-based kernel $\mathcal{K}^{EPI2}$ that builds inter-domain connections for the long-neglected sub-spaces $(u, y)$ and $(v, x)$, which is formulated as

\begin{equation}
    \mathcal{K}^{EPI2}(X)=F^{v,x}(F^{u,y}(X)).
\end{equation}

Fig. \ref{fig:Kernels} (b) and (c) illustrate these two EPI-based kernels.

\subsubsection{Combination Kernels}
The aforementioned kernels model the LF correlation in different ways as $\mathcal{K}^{SAS}$ is intra-domain, and $\mathcal{K}^{EPI1}$ and $\mathcal{K}^{EPI2}$ are inter-domain. We further combine the merits of intra-domain and inter-domain connections into two new kernels, namely $\mathcal{K}^{\alpha}$ and $\mathcal{K}^{\beta}$, which are illustrated in Fig. \ref{fig:Kernels} (d) and (e). Essentially, $\mathcal{K}^{\alpha}$ is $\mathcal{K}^{SAS}$ plus $\mathcal{K}^{EPI1}$, and $\mathcal{K}^{\beta}$ is $\mathcal{K}^{SAS}$ plus $\mathcal{K}^{EPI2}$:
\begin{equation}
    \mathcal{K}^{\alpha}(X)=F^{v,y}(F^{u,x}(F^{u,v}(F^{x,y}(X))))
\end{equation}
\begin{equation}
    \mathcal{K}^{\beta}(X) =F^{v,x}(F^{u,y}(F^{u,v}(F^{x,y}(X)))).
\end{equation}

Moreover, we combine the two EPI-based kernels, $\mathcal{K}^{EPI1}$ and $\mathcal{K}^{EPI2}$, to form $\mathcal{K}^{EPI3}$ defined as
\begin{equation}
    \mathcal{K}^{EPI3}(X)=F^{v,x}(F^{u,y}(F^{v,y}(F^{u,x}(X)))).
\end{equation}
$\mathcal{K}^{EPI3}$ can utilize all the possible inter-domain connections, and is illustrated in Fig. \ref{fig:Kernels} (f).

Finally, we exhaust all the connections to construct a full kernel $\mathcal{K}^{\gamma}$ as shown in Fig. \ref{fig:Kernels} (g), which is formulated as
\begin{equation}
    \mathcal{K}^{\gamma}(X)=F^{v,x}(F^{u,y}(F^{v,y}(F^{u,x}(F^{u,v}(F^{x,y}(X)))))).
\end{equation}

Compared with the SAS-based and EPI-based kernels, the hybridization of them can reinforce the spatio-angular representation with  connection diversity to improve the LFSR quality, which will be verified in Section \ref{section:Experiments}.

\subsubsection{Duplication Kernels}
Last but not least, we propose two duplication kernels, $\mathcal{K}^{dup1}$ and $\mathcal{K}^{dup2}$, by duplicating connections of $\mathcal{K}^{SAS}$. In $\mathcal{K}^{dup1}$, the $(x, y)$ connection is duplicated while the $(u, v)$ connection remains single. Likewise, in $\mathcal{K}^{dup2}$, the duplication is applied to the $(u, v)$ connection, and the $(x, y)$ connection remains the same. These two kernels are formulated as
\begin{equation}
    \mathcal{K}^{dup1}(X)=F^{u,v}(F^{x,y}_n(\dotsb(F^{x,y}_2(F^{x,y}_1(X)))))
\end{equation}
\begin{equation}
    \mathcal{K}^{dup2}(X)=F^{u,v}_n(\dotsb(F^{u,v}_2(F^{u,v}_1(F^{x,y}(X)))))
\end{equation}
where $F^{x,y}_i$ and $F^{u,v}_i$ indicate the $i$-th $F^{x,y}$ and $F^{u,v}$, and $n$ is the number of duplication. In Fig. \ref{fig:Kernels} (h) and (i), these two kernels with $n=3$ are illustrated respectively. These two kernels aim to extract deeper features in a specific sub-space by increasing the connection number one-sidedly.

So far, we have defined a wide range of decomposition kernels. The next subsection will introduce our proposed network that performs LFSR with these kernels.

\subsection{Network Architecture} \label{section:Network Architecture}

Based on the new decomposition kernels, we propose the Decomposition Kernel Network (DKNet) for LFSR as illustrated in Fig. \ref{fig:Network}. Following the pipeline of Yeung \textit{et al.} \cite{yeungSAS_LFSR2019}, the network consists of two main stages, namely feature extraction and image generation. In the feature extraction stage, the input LR image is first fed into a spatial convolutional layer $F^{x,y}$ to extract a preliminary spatial feature, then passed into a series of the proposed decomposition kernels $\{\mathcal{K}_1,\mathcal{K}_2,\dots,\mathcal{K}_L\}$ to extract deeper features. The number of decomposition kernels is denoted as $L$. Finally, the extracted features are aggregated and the feature channels are reduced in the image generation stage to generate the final HR images.

Besides stacking decomposition kernels in feature extraction, we improve the network with three major modifications.

Firstly, to overcome the gradient vanishment problem caused by the very deep structure, DKNet is equipped with dense connections, which have been successfully used for various computer vision tasks, \textit{e.g.} image classification \cite{huangDenseNet_CVPR2017} and SISR \cite{tongSRDenseNet_ICCV2017, zhangRDN_CVPR2018}. In DKNet, the dense connections can reinforce the information flow through the network by enabling later layers to access preceding layers. As a result, it supplies long-range information and aggregate hierarchical features to support the image generation stage. Therefore, the original feature extraction stage originally defined as
\begin{equation}
    \mathcal{K}_i = 
    \begin{cases}
        F(\mathcal{K}_{i-1}) & \text{if $i>1$}; \\
        F(X) & \text{if $i=1$} \\
    \end{cases}
\end{equation}
is consequently revised into
\begin{equation}
    \mathcal{K}_i = F([\mathcal{K}_{i-1}, \mathcal{K}_{i-2}, \dots, \mathcal{K}_1])
    \label{eq:KernelRevised1}
\end{equation}
where $[\cdot]$ denotes the concatenation operation.

Secondly, while Yeung \textit{et al.} \cite{yeungSAS_LFSR2019} utilized a residual connection \cite{heResNet_CVPR2016} to connect the input SAIs to the final image generation stage, we treat the input SAIs as a raw feature to be appended to the subsequent decomposition kernels by raw image connections, which is a special case of dense connections. This design has been adopted for video processing \cite{caballeroVESPCN_CVPR2017, ilgFlowNet2_CVPR2017} where raw frames are served as complementary information for motion estimation. Compared with the residual connection, which merely allows calculation of residue between the input and the output SAIs, in our case, the raw image connections engage the input in the calculation of all decomposition kernels, compensating the hierarchical extraction process with raw image information. Thus, Eq. \ref{eq:KernelRevised1} is revised to
\begin{equation}
    \mathcal{K}_i = F([\mathcal{K}_{i-1}, \mathcal{K}_{i-2}, \dots, \mathcal{K}_1, X]).
    \label{eq:KernelRevised2}
\end{equation}

\newcommand{\rd}[1]{\textbf{\textcolor{red}{#1}}}
\newcommand{\bl}[1]{\textbf{\textcolor{blue}{#1}}}

\begin{table*}[t!]
    \caption{Quantitative comparison (PSNR/SSIM) with state-of-the-art methods. The first and second best results are in \rd{red} and \bl{blue}.}
    \label{tab:PerfCompare}
    \centering
    \begin{tabu}{|l|c|c|c|c|c|c|}
    \hline
    Method                                                  & Scale        & Miscellaneous \cite{StanfordLytro} & Occlusions \cite{StanfordLytro} & Reflective \cite{StanfordLytro} & Kalantari \textit{et al.} \cite{kalantari2016learning} & EPFL \cite{rerabekEPFL2016} \\ 
    (\#Samples)                                             &              & (57)                               & (39)                            & (26)                            & (21)                                                   & (118) \\ \hline\hline
    
    SAN \cite{daiSAN_CVPR2019}                              & $2\times$    & 36.84/0.9556                       & 37.41/0.9721                    & 38.01/0.9650                    & 39.41/0.9804                                          & 39.39/0.9714  \\
    resLF \cite{zhangResLF_CVPR2019}                        & $2\times$    & 37.99/0.9690                       & 37.64/0.9773                    & 38.09/0.9643                    & 39.35/0.9794                                          & 39.54/0.9716  \\
    HDDRNet \cite{mengHDDRNet_AAAI2020}                     & $2\times$    & 34.86/0.9337                       & 34.61/0.9405                    & 35.69/0.9356                    & 36.77/0.9517                                          & 36.50/0.9404  \\
    LFSSR-ATO \cite{jinLFSSRATO_2020}                       & $2\times$    & 39.89/0.9734                       & \rd{40.39}/0.9786               & \rd{40.43}/0.9698               & \bl{42.61}/0.9825                                     & 41.63/0.9742  \\
    LF-InterNet \cite{wangLfInterNet_ECCV2020}              & $2\times$    & 39.42/0.9532                       & \bl{40.12}/0.9723               & 40.01/0.9612                    & 42.56/0.9800                                          & 41.79/0.9744  \\
    Yeung \textit{et al.} (SAS) \cite{yeungSAS_LFSR2019}    & $2\times$    & 40.37/0.9775                       & 39.81/0.9789                    & 39.98/\bl{0.9811}               & 42.41/0.9890                                          & 41.61/0.9782  \\
    Yeung \textit{et al.} (4D) \cite{yeungSAS_LFSR2019}     & $2\times$    & \bl{40.67}/\bl{0.9785}             & 40.01/\bl{0.9801}               & 40.11/\rd{0.9834}               & 42.60/\bl{0.9912}                                     & \bl{41.80}/\bl{0.9823}  \\
    DKNet (Ours)                                            & $2\times$    & \rd{42.02}/\rd{0.9928}             & 40.05/\rd{0.9836}               & \bl{40.26}/\rd{0.9834}          & \rd{42.89}/\rd{0.9932}                                & \rd{42.04}/\rd{0.9893}  \\ \hline\hline
    
    SAN \cite{daiSAN_CVPR2019}                              & $3\times$    & 33.02/0.9003                       & 31.12/0.8801                    & 31.42/0.8907                    & 31.63/0.8922                                          & 31.55/0.8941  \\
    HDDRNet \cite{mengHDDRNet_AAAI2020}                     & $3\times$    & 32.79/0.8950                       & 32.20/0.8832                    & 33.58/0.8979                    & 35.13/0.9254                                          & 34.74/0.9090  \\
    Yeung \textit{et al.} (SAS) \cite{yeungSAS_LFSR2019}    & $3\times$    & \bl{37.00}/\bl{0.9556}             & \bl{36.24}/\bl{0.9636}          & \bl{37.39}/\bl{0.9735}          & \bl{39.73}/\bl{0.9866}                                & \bl{38.93}/\bl{0.9795}  \\
    Yeung \textit{et al.} (4D) \cite{yeungSAS_LFSR2019}     & $3\times$    & 36.64/0.9511                       & 35.46/0.9578                    & 37.01/0.9699                    & 39.33/0.9854                                          & 38.64/0.9780  \\
    DKNet (Ours)                                            & $3\times$    & \rd{37.83}/\rd{0.9831}             & \rd{36.87}/\rd{0.9649}          & \rd{37.71}/\rd{0.9758}          & \rd{40.28}/\rd{0.9888}                                & \rd{39.78}/\rd{0.9828}  \\ \hline\hline
    
    SAN \cite{daiSAN_CVPR2019}                              & $4\times$    & 30.83/0.8491                       & 29.89/0.8112                    & 29.91/0.8110                    & 30.16/0.8439                                          & 30.01/0.8361  \\
    resLF \cite{zhangResLF_CVPR2019}                        & $4\times$    & 33.33/0.9610                       & 31.31/0.8854                    & 33.05/0.8977                    & 34.84/0.9390                                          & 34.58/0.9151  \\
    HDDRNet \cite{mengHDDRNet_AAAI2020}                     & $4\times$    & 30.58/0.8412                       & 29.82/0.8119                    & 31.22/0.8508                    & 33.02/0.8944                                          & 32.51/0.8598  \\
    LFSSR-ATO \cite{jinLFSSRATO_2020}                       & $4\times$    & 33.41/0.9068                       & 32.72/0.8849                    & \bl{34.82}/0.9119               & 36.43/0.9443                                          & 35.93/0.9227  \\
    LF-InterNet \cite{wangLfInterNet_ECCV2020}              & $4\times$    & 33.42/0.9532                       & 32.55/0.9342                    & 34.46/0.9556                    & 36.56/0.9742                                          & \bl{36.19}/0.9631  \\
    Yeung \textit{et al.} (SAS) \cite{yeungSAS_LFSR2019}    & $4\times$    & 33.59/0.9602                       & 32.36/0.9321                    & 34.50/0.9531                    & 36.33/0.9727                                          & 35.93/0.9618  \\
    Yeung \textit{et al.} (4D) \cite{yeungSAS_LFSR2019}     & $4\times$    & \bl{34.27}/\bl{0.9611}             & \bl{32.98}/\bl{0.9373}          & \rd{34.97}/\bl{0.9560}          & \bl{36.42}/\bl{0.9750}                                & 36.03/\bl{0.9651}  \\
    DKNet (Ours)                                            & $4\times$    & \rd{36.07}/\rd{0.9752}             & \rd{33.12}/\rd{0.9408}          & 34.50/\rd{0.9600}               & \rd{36.73}/\rd{0.9783}                                & \rd{36.29}/\rd{0.9675}  \\ \hline
    \end{tabu}
\end{table*}

Lastly, we adopt a pixel shuffler \cite{shiESPCN_CVPR2016} instead of the transposed convolution in \cite{yeungSAS_LFSR2019} for upsampling the spatial resolution in image generation. This approach can avoid artifacts and generate more complex details \cite{shiESPCN_CVPR2016}. Another remarkable benefit is that the pixel shuffler allows upsampling to an arbitrary scale without any major modification on the network. In \cite{yeungSAS_LFSR2019}, the network conducted $4 \times$ super-resolution by duplicating the network of $2 \times$ scale, thus incurring twice the cost in computation and memory.

Before the pixel shuffler, similar to \cite{yeungSAS_LFSR2019}, feature reduction is conducted to aggregate the dense features and reduce the angular space to $(1 \times 1)$ to allow the pixel shuffler to perform. When processing an $(8 \times 8)$ LF image, two convolutions are exploited to reduce the angular space to $(4 \times 4)$ and then $(1 \times 1)$. Table \ref{tab:Archi} gives more details of the network architecture.

\subsection{Loss Functions} \label{section:Loss Functions}
To train the networks for LFSR, the existing methods typically adopted pixel-wise loss functions to minimize the pixel-wise error and achieve higher PSNR. Among them, the mean square error (MSE) is the most widely used option \cite{yeungSAS_LFSR2019, yoon2017LFCNN, gul2018spatial} which is defined as:
\begin{equation}\label{eq:pixel loss}
    \mathcal{L}_{MSE}(Y, \hat{Y}) = \frac{1}{UVrWrHC}  \|Y - \hat{Y} \|^2.
\end{equation}
where $C$ is the number of channels which is usually 3 in the RGB color space. Although pixel-wise loss functions achieve higher performance under pixel-wise metrics, they don't specialize in capturing perceptual differences in a broad context and tend to produce over-smoothed images, especially in areas with rich details. Instead of solely relying on the pixel-wise loss, inspired by the success of VGG-based loss functions in SISR \cite{johnsonVGGLoss_ECCV2016}, we propose a novel LFVGG loss to pursue a similar improvement of LFSR. Different from the single-image occasion, in LFSR, the correlated LF structure and the extra two dimensions should be taken into consideration. Specifically, the LFVGG loss minimizes the SAI-wise error in the feature space of the pre-trained VGG-19 network \cite{simonyanVGG_2015}. Therefore, the LFVGG loss is formulated as 
\begin{equation}\label{eq:feature loss}
    \mathcal{L}_{LFVGG}(Y, \hat{Y}) = \frac{1}{UV W_l H_l C_l} \|\phi_l (Y)- \phi_l (\hat{Y}) \|^2
\end{equation}
where $\phi_l(\cdot)$ is the output of layer $l$ of the VGG-19 network performed on the $U \times V$ SAIs individually. The selected convolutional layer $l$ determines the spatial size and the channel number $(W_l, H_l, C_l)$ of the tensor size $(U, V, W_l, H_l, C_l)$. The final loss function is the weighted sum of Eq. \ref{eq:pixel loss} and \ref{eq:feature loss}:
\begin{equation}\label{eq:final loss}
    \mathcal{L}(Y, \hat{Y})  = \mathcal{L}_{MSE}(Y, \hat{Y}) + \lambda \mathcal{L}_{LFVGG}(Y, \hat{Y}).
\end{equation}

In \cite{johnsonVGGLoss_ECCV2016}, due to the loosen constraint of exact pixel locations, the VGG-based loss function often led to a lower PSNR, however, their network was guided by the loss to produce pleasing visual results with enhanced textures. In our work, similar effects are brought to LF images, but we will exploit an indirect metric to evaluate the visual improvement objectively in Section \ref{section:StudyOfFeatureLoss}.
\section{Experiments} \label{section:Experiments}
\subsection{Implementation and Evaluation Details}
The proposed DKNet is implemented with the deep learning library PyTorch \cite{PyTorch}. The source code and models will be released publicly at \href{https://huzexi.github.io}{https://huzexi.github.io}.

We follow the evaluation protocol of Yeung \textit{et al.} \cite{yeungSAS_LFSR2019} to conduct the experiments. The training set consists of 130 LF samples from Stanford Lytro Light Field Archive \cite{StanfordLytro} and Kalantari \textit{et al.} \cite{kalantari2016learning}. All these LF images are captured by Lytro Illum cameras and come with a size of $14 \times 14 \times 540 \times 376 \times 3$. Only the middle $8\times 8 \times 510 \times 346 \times 3$ is used in the experiments discarding 3 marginal SAIs on each boundary and 15 marginal pixels of each SAI due to their poor quality. A bilinear interpolation is used for down-sampling. We train the model with an Adam optimizer \cite{kingmaAdam_ICLR2015} with a learning rate of $10^{-4}$. The batch size is 2 and the patch spatial size is $(32 \times 32)$.

\begin{figure*}[t!]
    \centering
    \resizebox{\textwidth}{!}{
    \tabcolsep=0.02cm
    \renewcommand{\arraystretch}{0.4}
    \begin{tabular}{cccccccccccc}
        \multicolumn{1}{l}{}
        & \multicolumn{2}{c}{LR}
        & \multicolumn{2}{c}{SAN \cite{daiSAN_CVPR2019}}
        & \multicolumn{2}{c}{Yeung \textit{et al.} (4D) \cite{yeungSAS_LFSR2019}}
        & \multicolumn{2}{c}{DKNet (Ours)}
        & \multicolumn{2}{c}{Ground-truth}
        \\
        
        \multicolumn{1}{c}{\rotatebox{90}{\hspace{12pt}{\textit{general\_1}}}}
        & \multicolumn{2}{l}{\includegraphics[width=0.19\textwidth, height=0.130\textwidth]{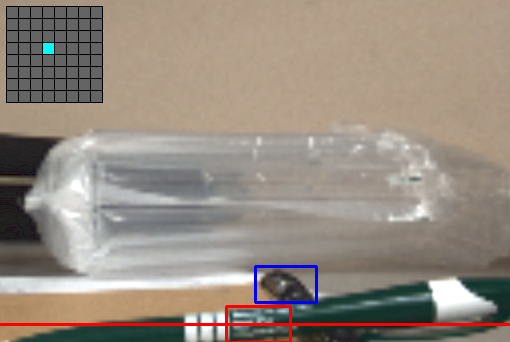}}
        & \multicolumn{2}{l}{\includegraphics[width=0.19\textwidth, height=0.130\textwidth]{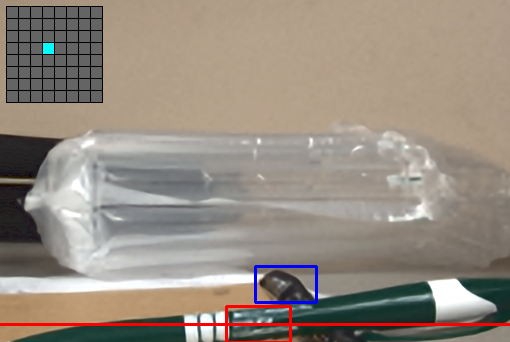}}
        & \multicolumn{2}{l}{\includegraphics[width=0.19\textwidth, height=0.130\textwidth]{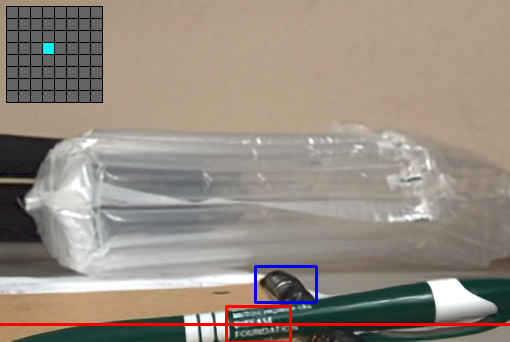}}
        & \multicolumn{2}{l}{\includegraphics[width=0.19\textwidth, height=0.130\textwidth]{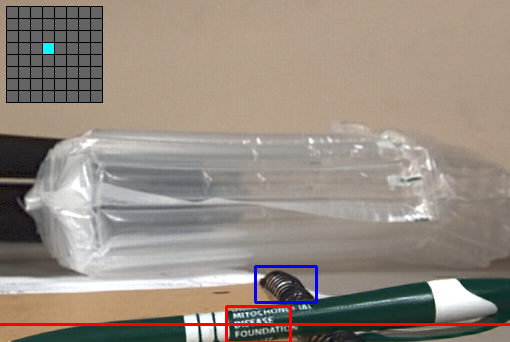}}
        & \multicolumn{2}{l}{\includegraphics[width=0.19\textwidth, height=0.130\textwidth]{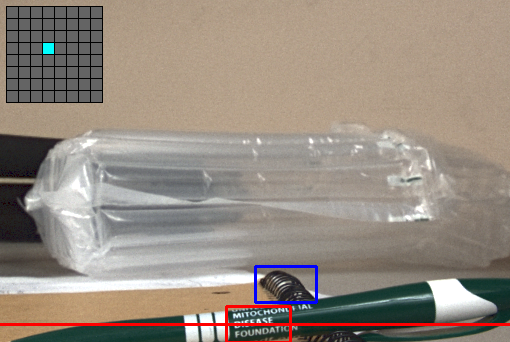}}
        &
        \\
        
        & \multicolumn{1}{l}{\includegraphics[width=0.09\textwidth, height=0.052\textwidth,cfbox=red 1pt 0pt] {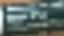}}
        & \multicolumn{1}{r}{\includegraphics[width=0.09\textwidth, height=0.052\textwidth,cfbox=blue 1pt 0pt]{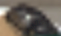}}
        & \multicolumn{1}{l}{\includegraphics[width=0.09\textwidth, height=0.052\textwidth,cfbox=red 1pt 0pt] {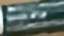}}
        & \multicolumn{1}{r}{\includegraphics[width=0.09\textwidth, height=0.052\textwidth,cfbox=blue 1pt 0pt]{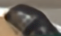}}
        & \multicolumn{1}{l}{\includegraphics[width=0.09\textwidth, height=0.052\textwidth,cfbox=red 1pt 0pt] {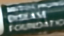}}
        & \multicolumn{1}{r}{\includegraphics[width=0.09\textwidth, height=0.052\textwidth,cfbox=blue 1pt 0pt]{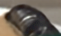}}
        & \multicolumn{1}{l}{\includegraphics[width=0.09\textwidth, height=0.052\textwidth,cfbox=red 1pt 0pt] {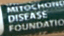}}
        & \multicolumn{1}{r}{\includegraphics[width=0.09\textwidth, height=0.052\textwidth,cfbox=blue 1pt 0pt]{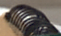}}
        & \multicolumn{1}{l}{\includegraphics[width=0.09\textwidth, height=0.052\textwidth,cfbox=red 1pt 0pt] {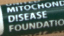}}
        & \multicolumn{1}{r}{\includegraphics[width=0.09\textwidth, height=0.052\textwidth,cfbox=blue 1pt 0pt]{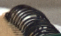}}
        &
        \\

        \multicolumn{1}{c}{}
        & \multicolumn{2}{l}{\includegraphics[width=0.19\textwidth, height=0.016\textheight]{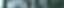}}
        & \multicolumn{2}{l}{\includegraphics[width=0.19\textwidth, height=0.016\textheight]{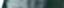}}
        & \multicolumn{2}{l}{\includegraphics[width=0.19\textwidth, height=0.016\textheight]{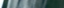}}
        & \multicolumn{2}{l}{\includegraphics[width=0.19\textwidth, height=0.016\textheight]{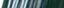}}
        & \multicolumn{2}{l}{\includegraphics[width=0.19\textwidth, height=0.016\textheight]{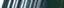}}
        &
        \\

        \multicolumn{1}{c}{}
        & \multicolumn{2}{l}{\includegraphics[width=0.19\textwidth, height=0.016\textheight]{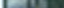}}
        & \multicolumn{2}{l}{\includegraphics[width=0.19\textwidth, height=0.016\textheight]{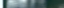}}
        & \multicolumn{2}{l}{\includegraphics[width=0.19\textwidth, height=0.016\textheight]{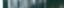}}
        & \multicolumn{2}{l}{\includegraphics[width=0.19\textwidth, height=0.016\textheight]{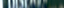}}
        & \multicolumn{2}{l}{\includegraphics[width=0.19\textwidth, height=0.016\textheight]{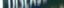}}
        &
        \\

        \multicolumn{1}{c}{\vspace{0.15cm}}
        & \multicolumn{2}{c}{32.32/0.8959}
        & \multicolumn{2}{c}{34.42/0.9658}
        & \multicolumn{2}{c}{36.90/0.9777}
        & \multicolumn{2}{c}{38.48/0.9807}
        & \multicolumn{2}{c}{}
        \\

        \multicolumn{1}{c}{\rotatebox{90}{\hspace{12pt}{\textit{general\_23}}}}
        & \multicolumn{2}{l}{\includegraphics[width=0.19\textwidth, height=0.130\textwidth]{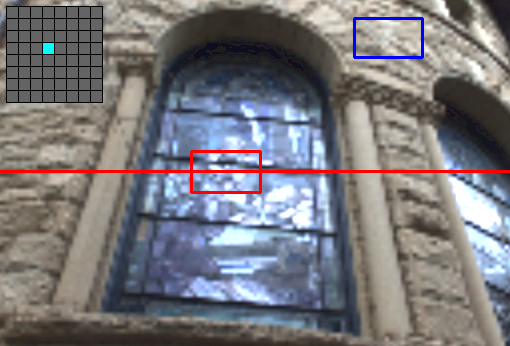}}
        & \multicolumn{2}{l}{\includegraphics[width=0.19\textwidth, height=0.130\textwidth]{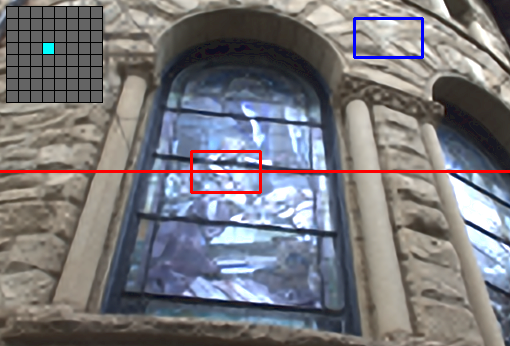}}
        & \multicolumn{2}{l}{\includegraphics[width=0.19\textwidth, height=0.130\textwidth]{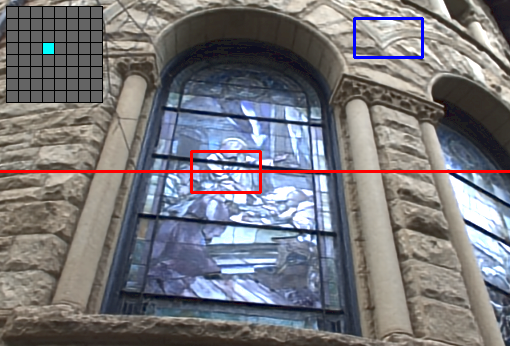}}
        & \multicolumn{2}{l}{\includegraphics[width=0.19\textwidth, height=0.130\textwidth]{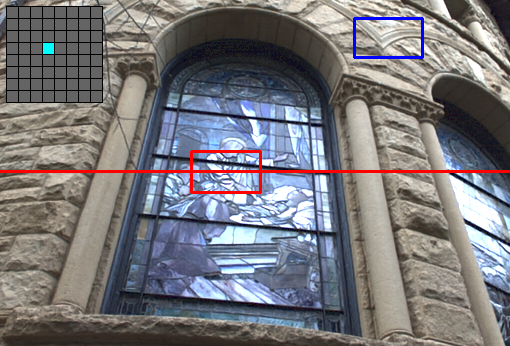}}
        & \multicolumn{2}{l}{\includegraphics[width=0.19\textwidth, height=0.130\textwidth]{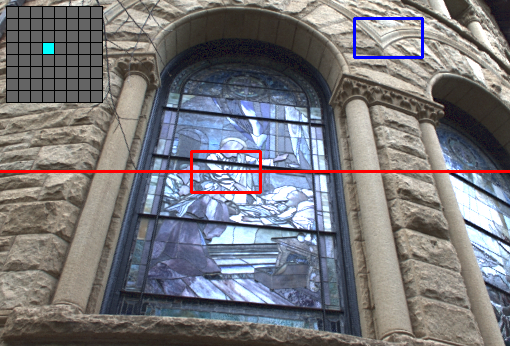}}
        &
        \\
        
        & \multicolumn{1}{l}{\includegraphics[width=0.09\textwidth, height=0.052\textwidth,cfbox=red 1pt 0pt] {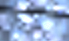}}
        & \multicolumn{1}{r}{\includegraphics[width=0.09\textwidth, height=0.052\textwidth,cfbox=blue 1pt 0pt]{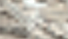}}
        & \multicolumn{1}{l}{\includegraphics[width=0.09\textwidth, height=0.052\textwidth,cfbox=red 1pt 0pt] {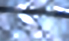}}
        & \multicolumn{1}{r}{\includegraphics[width=0.09\textwidth, height=0.052\textwidth,cfbox=blue 1pt 0pt]{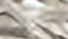}}
        & \multicolumn{1}{l}{\includegraphics[width=0.09\textwidth, height=0.052\textwidth,cfbox=red 1pt 0pt] {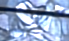}}
        & \multicolumn{1}{r}{\includegraphics[width=0.09\textwidth, height=0.052\textwidth,cfbox=blue 1pt 0pt]{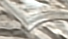}}
        & \multicolumn{1}{l}{\includegraphics[width=0.09\textwidth, height=0.052\textwidth,cfbox=red 1pt 0pt] {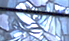}}
        & \multicolumn{1}{r}{\includegraphics[width=0.09\textwidth, height=0.052\textwidth,cfbox=blue 1pt 0pt]{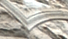}}
        & \multicolumn{1}{l}{\includegraphics[width=0.09\textwidth, height=0.052\textwidth,cfbox=red 1pt 0pt] {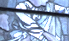}}
        & \multicolumn{1}{r}{\includegraphics[width=0.09\textwidth, height=0.052\textwidth,cfbox=blue 1pt 0pt]{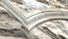}}
        &
        \\

        \multicolumn{1}{c}{}
        & \multicolumn{2}{l}{\includegraphics[width=0.19\textwidth, height=0.016\textheight]{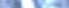}}
        & \multicolumn{2}{l}{\includegraphics[width=0.19\textwidth, height=0.016\textheight]{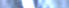}}
        & \multicolumn{2}{l}{\includegraphics[width=0.19\textwidth, height=0.016\textheight]{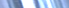}}
        & \multicolumn{2}{l}{\includegraphics[width=0.19\textwidth, height=0.016\textheight]{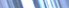}}
        & \multicolumn{2}{l}{\includegraphics[width=0.19\textwidth, height=0.016\textheight]{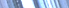}}
        &
        \\

        \multicolumn{1}{c}{}
        & \multicolumn{2}{l}{\includegraphics[width=0.19\textwidth, height=0.016\textheight]{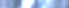}}
        & \multicolumn{2}{l}{\includegraphics[width=0.19\textwidth, height=0.016\textheight]{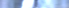}}
        & \multicolumn{2}{l}{\includegraphics[width=0.19\textwidth, height=0.016\textheight]{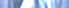}}
        & \multicolumn{2}{l}{\includegraphics[width=0.19\textwidth, height=0.016\textheight]{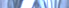}}
        & \multicolumn{2}{l}{\includegraphics[width=0.19\textwidth, height=0.016\textheight]{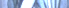}}
        &
        \\

        \multicolumn{1}{c}{\vspace{0.15cm}}
        & \multicolumn{2}{c}{24.27/0.6669}
        & \multicolumn{2}{c}{25.67/0.8367}
        & \multicolumn{2}{c}{30.43/0.9466}
        & \multicolumn{2}{c}{34.64/0.9789}
        & \multicolumn{2}{c}{}
        \\

        \multicolumn{1}{c}{\rotatebox{90}{\hspace{12pt}{\textit{general\_26}}}}
        & \multicolumn{2}{l}{\includegraphics[width=0.19\textwidth, height=0.130\textwidth]{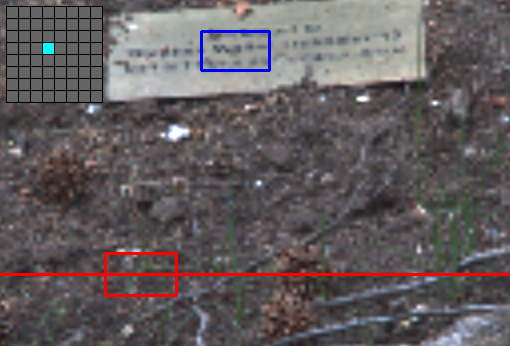}}
        & \multicolumn{2}{l}{\includegraphics[width=0.19\textwidth, height=0.130\textwidth]{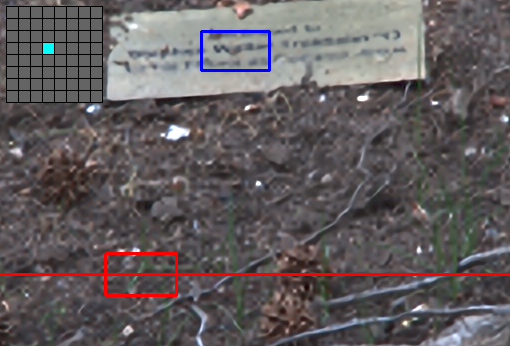}}
        & \multicolumn{2}{l}{\includegraphics[width=0.19\textwidth, height=0.130\textwidth]{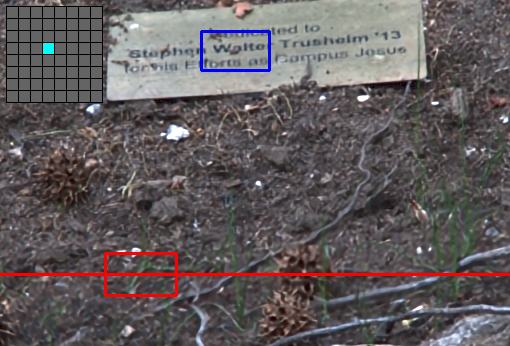}}
        & \multicolumn{2}{l}{\includegraphics[width=0.19\textwidth, height=0.130\textwidth]{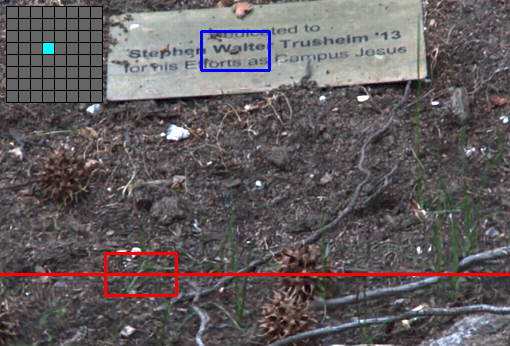}}
        & \multicolumn{2}{l}{\includegraphics[width=0.19\textwidth, height=0.130\textwidth]{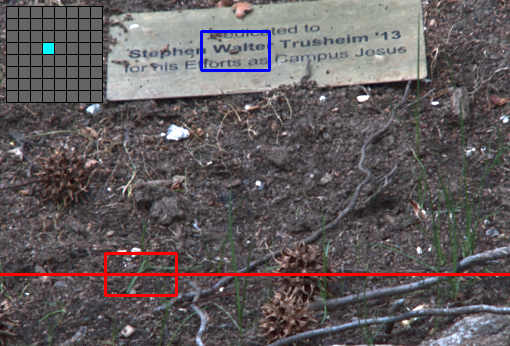}}
        &
        \\
        
        & \multicolumn{1}{l}{\includegraphics[width=0.09\textwidth, height=0.052\textwidth,cfbox=red 1pt 0pt] {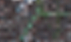}}
        & \multicolumn{1}{r}{\includegraphics[width=0.09\textwidth, height=0.052\textwidth,cfbox=blue 1pt 0pt]{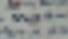}}
        & \multicolumn{1}{l}{\includegraphics[width=0.09\textwidth, height=0.052\textwidth,cfbox=red 1pt 0pt] {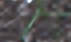}}
        & \multicolumn{1}{r}{\includegraphics[width=0.09\textwidth, height=0.052\textwidth,cfbox=blue 1pt 0pt]{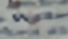}}
        & \multicolumn{1}{l}{\includegraphics[width=0.09\textwidth, height=0.052\textwidth,cfbox=red 1pt 0pt] {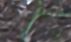}}
        & \multicolumn{1}{r}{\includegraphics[width=0.09\textwidth, height=0.052\textwidth,cfbox=blue 1pt 0pt]{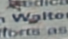}}
        & \multicolumn{1}{l}{\includegraphics[width=0.09\textwidth, height=0.052\textwidth,cfbox=red 1pt 0pt] {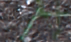}}
        & \multicolumn{1}{r}{\includegraphics[width=0.09\textwidth, height=0.052\textwidth,cfbox=blue 1pt 0pt]{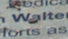}}
        & \multicolumn{1}{l}{\includegraphics[width=0.09\textwidth, height=0.052\textwidth,cfbox=red 1pt 0pt] {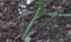}}
        & \multicolumn{1}{r}{\includegraphics[width=0.09\textwidth, height=0.052\textwidth,cfbox=blue 1pt 0pt]{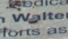}}
        &
        \\

        \multicolumn{1}{c}{}
        & \multicolumn{2}{l}{\includegraphics[width=0.19\textwidth, height=0.016\textheight]{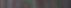}}
        & \multicolumn{2}{l}{\includegraphics[width=0.19\textwidth, height=0.016\textheight]{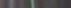}}
        & \multicolumn{2}{l}{\includegraphics[width=0.19\textwidth, height=0.016\textheight]{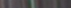}}
        & \multicolumn{2}{l}{\includegraphics[width=0.19\textwidth, height=0.016\textheight]{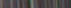}}
        & \multicolumn{2}{l}{\includegraphics[width=0.19\textwidth, height=0.016\textheight]{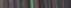}}
        &
        \\

        \multicolumn{1}{c}{}
        & \multicolumn{2}{l}{\includegraphics[width=0.19\textwidth, height=0.016\textheight]{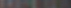}}
        & \multicolumn{2}{l}{\includegraphics[width=0.19\textwidth, height=0.016\textheight]{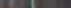}}
        & \multicolumn{2}{l}{\includegraphics[width=0.19\textwidth, height=0.016\textheight]{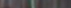}}
        & \multicolumn{2}{l}{\includegraphics[width=0.19\textwidth, height=0.016\textheight]{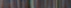}}
        & \multicolumn{2}{l}{\includegraphics[width=0.19\textwidth, height=0.016\textheight]{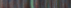}}
        &
        \\

        \multicolumn{1}{c}{\vspace{0.15cm}}
        & \multicolumn{2}{c}{26.69/0.6184}
        & \multicolumn{2}{c}{27.41/0.8121}
        & \multicolumn{2}{c}{31.46/0.9257}
        & \multicolumn{2}{c}{33.42/0.9587}
        & \multicolumn{2}{c}{}
        \\
        
    \end{tabular}
    }
    \caption{Visualization of selective methods' results. Three samples are presented in three rows respectively. In each row, a SAI, two zoomed-in views (red and blue boxes), two EPIs (along the red line and two angular dimensions in the red box), and PSNR/SSIM are given.}
    \label{fig:Qual1}
\end{figure*}

\subsection{Comparison with State-of-the-art methods} \label{section:ComparisonPixelWise}

\subsubsection{Quantitative Comparison}
To demonstrate the superiority of DKNet, we compare our proposed method with the state-of-the-art methods, namely resLF \cite{zhangResLF_CVPR2019}, HDDRNet \cite{mengHDDRNet_AAAI2020}, LFSSR-ATO \cite{jinLFSSRATO_2020}, LF-InterNet \cite{wangLfInterNet_ECCV2020} and Yeung \textit{et al.} \cite{yeungSAS_LFSR2019} (the SAS and 4D versions). The most recent SISR method, SAN \cite{daiSAN_CVPR2019} is also tested to demonstrate the merit of utilizing correlation information of SAIs. To achieve higher performance under the conventional PSNR and SSIM metrics, DKNet is trained with the pixel-wise loss function of Eq. \ref{eq:pixel loss}. In terms of hyperparameters, we empirically set the number of decomposition kernels $L$ to 18, and the convolution channels to 32. $\mathcal{K}^{\gamma}$ is set as the decomposition kernels to achieve the best performance as it comes with the most diverse connections. The evaluation is conducted on an extended range of datasets including: 1) Miscellaneous, Occlusions and Reflective Categories of Stanford Lytro Light Field Archive \cite{StanfordLytro}; 2) The test set of Kalantari \textit{et al.} \cite{kalantari2016learning}; and 3) EPFL dataset \cite{rerabekEPFL2016}. The samples that have been used in the training set are removed to ensure a fair evaluation. We evaluate the methods in three LFSR scales including $2 \times$, $3 \times$ and $4 \times$, and the results are shown in Table \ref{tab:PerfCompare}. Due to the $3\times$ scale models of resLF \cite{zhangResLF_CVPR2019}, LFSSR-ATO \cite{jinLFSSRATO_2020} and LF-InterNet \cite{wangLfInterNet_ECCV2020} are not publicly available, they are absent in $3 \times$ scale.

The results show that all the LFSR methods, except HDDRNet \cite{mengHDDRNet_AAAI2020}, surpass the SISR method, SAN \cite{daiSAN_CVPR2019}, to various extents, verifying the effectiveness of the correlation information in LFSR. When comparing the LFSR methods, DKNet outperforms the previous state-of-the-art works on most of the datasets. Noticeably, in the most challenging $4\times$ task, DKNet outperforms the second-best Yeung \textit{et al.} (4D) \cite{yeungSAS_LFSR2019} in Miscellaneous dataset by a significant margin of 1.80 dB PSNR. Regarding the other general datasets, DKNet still outperforms the second-best method by 0.14 dB on Occlusions dataset, 0.31 dB on Kalantari \textit{et al.} \cite{kalantari2016learning} and 0.10 dB on EPFL \cite{rerabekEPFL2016}. In Reflective dataset, which features the challenge of reflective surfaces like mirrors and water, DKNet shows comparable performance to the second-best method Yeung \textit{et al.} (4D) \cite{yeungSAS_LFSR2019} with a slightly lower PSNR but a higher SSIM. Overall, DKNet is still the leading model among all the tested methods under the SSIM metric across all the datasets by between 0.0024 to 0.0141. The difference implies the LF image super-solved by DKNet has a better quality in regards to luminance, contrast, and structure, which are more correlated to human visual perception than PSNR which measures the pixel-to-pixel difference \cite{luSSIM_AAAI2019}.

A similar trend can be observed in scale $2\times$ where DKNet outperforms the competitors on the five datasets under the SSIM metric and on the three datasets under the PSNR metric. When looking at scale $3\times$, DKNet consistently produces the best results under the two metrics on all of the datasets.

\begin{figure}[ht!]
    \centering
    \tabcolsep=0.03cm
    \renewcommand{\arraystretch}{0.25}
    \begin{tabu}{ccccc}
        &&
        \scriptsize Original &
        \scriptsize Yeung (4D) \cite{yeungSAS_LFSR2019}  &
        \scriptsize DKNet
        \\

        \multirow{3}{*}{\rotatebox{90}{\scriptsize\hspace{45pt}\textit{occlusion\_9}}} &
        \multirow{3}{*}[0.75cm]{\includegraphics[width=0.140\textwidth]{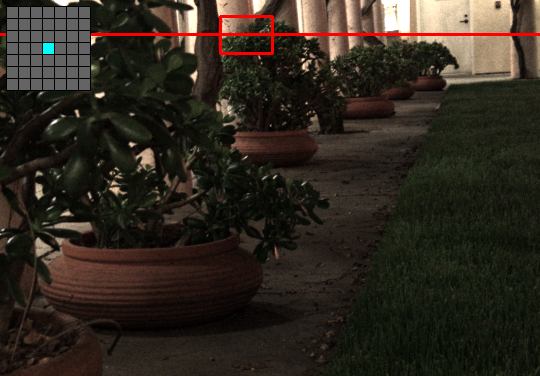}} &
        \includegraphics[width=0.100\textwidth, height=1.10cm] {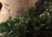} &
        \includegraphics[width=0.100\textwidth, height=1.10cm] {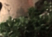} &
        \includegraphics[width=0.100\textwidth, height=1.10cm] {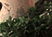} \\
        &&
        \includegraphics[width=0.100\textwidth, height=0.30cm] {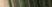} &
        \includegraphics[width=0.100\textwidth, height=0.30cm] {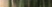} &
        \includegraphics[width=0.100\textwidth, height=0.30cm] {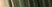} \\
        &&
        \includegraphics[width=0.100\textwidth, height=0.30cm] {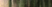} &
        \includegraphics[width=0.100\textwidth, height=0.30cm] {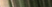} &
        \includegraphics[width=0.100\textwidth, height=0.30cm] {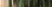} \\

        \vspace{0.00cm} \\

        \multirow{3}{*}{\rotatebox{90}{\scriptsize\hspace{42pt}\textit{occlusions\_42}}} &
        \multirow{3}{*}[0.75cm]{\includegraphics[width=0.140\textwidth]{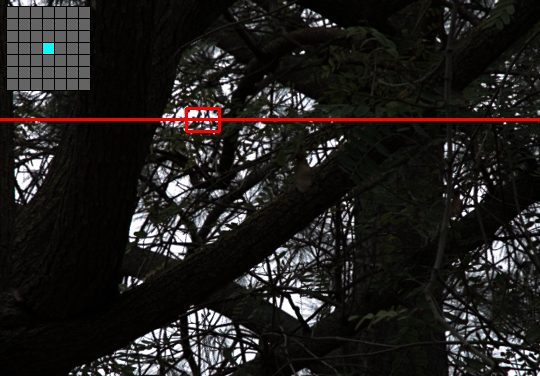}} &
        \includegraphics[width=0.100\textwidth, height=1.10cm] {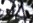} &
        \includegraphics[width=0.100\textwidth, height=1.10cm] {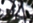} &
        \includegraphics[width=0.100\textwidth, height=1.10cm] {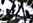} \\
        &&
        \includegraphics[width=0.100\textwidth, height=0.30cm] {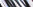} &
        \includegraphics[width=0.100\textwidth, height=0.30cm] {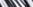} &
        \includegraphics[width=0.100\textwidth, height=0.30cm] {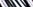} \\
        &&
        \includegraphics[width=0.100\textwidth, height=0.30cm] {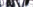} &
        \includegraphics[width=0.100\textwidth, height=0.30cm] {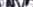} &
        \includegraphics[width=0.100\textwidth, height=0.30cm] {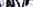} \\
        
        \vspace{0.00cm} \\

        \multirow{3}{*}{\rotatebox{90}{\scriptsize\hspace{42pt}\textit{reflective\_18}}} &
        \multirow{3}{*}[0.75cm]{\includegraphics[width=0.140\textwidth]{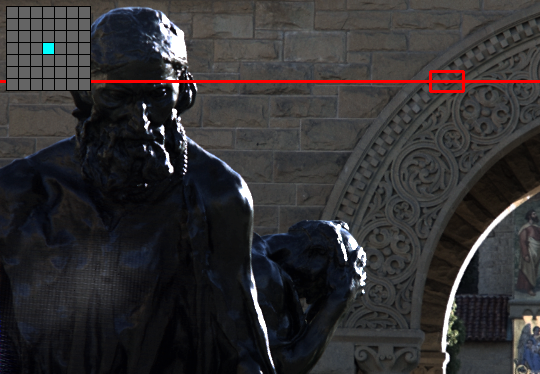}} &
        \includegraphics[width=0.100\textwidth, height=1.10cm] {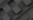} &
        \includegraphics[width=0.100\textwidth, height=1.10cm] {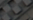} &
        \includegraphics[width=0.100\textwidth, height=1.10cm] {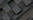} \\
        &&
        \includegraphics[width=0.100\textwidth, height=0.30cm] {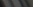} &
        \includegraphics[width=0.100\textwidth, height=0.30cm] {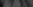} &
        \includegraphics[width=0.100\textwidth, height=0.30cm] {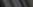} \\
        &&
        \includegraphics[width=0.100\textwidth, height=0.30cm] {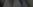} &
        \includegraphics[width=0.100\textwidth, height=0.30cm] {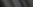} &
        \includegraphics[width=0.100\textwidth, height=0.30cm] {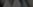} \\
        
        \vspace{0.00cm} \\
    \end{tabu}
    \caption{Visualization of further enhancement. Two EPIs are extracted along the red line and the two angular dimensions in the red box separately.}
    \label{fig:FurtherEnhancement}
\end{figure}

These results suggest DKNet's overall advantage of LFSR with better utilization of the correlation information and more powerful spatio-angular feature representation.

\subsubsection{Qualitative Comparison}
For a better understanding of the performance, some results of scale $4 \times$ are visualized in Fig. \ref{fig:Qual1}. DKNet is compared with SAN \cite{daiSAN_CVPR2019} and Yeung \textit{et al.} \cite{yeungSAS_LFSR2019} as the representatives of SISR and LFSR methods.

It can be observed that DKNet has produced better results with clearer and richer details, particularly for the text areas. For example, as shown in the red box in \textit{general\_1} and the blue box in \textit{general\_26}, DKNet is the only method that recovers the text to a readable level. Also, in \textit{general\_23}, the edges of the building and the patterns on the glass window are well recovered with sharpness and integrity by DKNet while SAN \cite{daiSAN_CVPR2019} and Yeung \textit{et al.} \cite{yeungSAS_LFSR2019} both output blurry images. Successful recovery of complicated details could also be observed in the blue box of \textit{general\_1} where the compact edges of the spiral bindings are more distinguishable in the result of DKNet. In \textit{general\_26}, while the other methods produce blurry images, DKNet produces a sharper image, especially at the border of the thin leaf and its soil background in the red box.

The EPIs of two angular directions in the red boxes are presented to demonstrate the improvement in these two sub-spaces. As depicted, DKNet could produce results that are closest to the ground-truth while Yeung \textit{et al.} \cite{yeungSAS_LFSR2019} can only recover the pattern to a limited extent. In \textit{general\_1}, the patterns are more distinguishable as the printed text is recovered. In \textit{general\_23}, the EPIs are not only sharper but also more informative with more edges in DKNet's second EPI. In \textit{general\_26}, the EPIs basically conform to the observation of SAIs as the LF images produced by DKNet are sharper near the leaf and informative in the soil background. SAN \cite{daiSAN_CVPR2019} is completely unable to make use of the correlation and generates blurry EPIs, and Yeung \textit{et al.} \cite{yeungSAS_LFSR2019} cannot make the best use of the sub-spaces and fails to recover as much information as DKNet does.

\begin{table*}[t]
    \caption{Study of decomposition kernels on $4 \times$ scale. The number of connections, PSNR and SSIM are given. The best item is underlined.}
    \label{tab:PerfKernel}
    \centering
    \resizebox{\textwidth}{!}{
        \begin{tabular}{|l||c|c|c|c|c|c|c|c|c|c|c|}
        \hline
        \rule{0pt}{8pt}Kernel    & $\mathcal{K}^{SAS}$    & $\mathcal{K}^{EPI1}$    & $\mathcal{K}^{EPI2}$    & $\mathcal{K}^{EPI3}$     & $\mathcal{K}^{\alpha}$    & $\mathcal{K}^{\beta}$     & $\mathcal{K}^{\gamma}$        & $\mathcal{K}^{Dup1-4}$        & $\mathcal{K}^{Dup2-4}$    & $\mathcal{K}^{Dup1-6}$    & $\mathcal{K}^{Dup2-6}$  \\ \hline
        Illustration        & 
        \begin{minipage}{1.0cm}
            \includegraphics[width=1.0\textwidth]{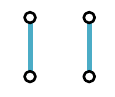}
        \end{minipage} & 
        \begin{minipage}{1.0cm}
            \includegraphics[width=1.0\textwidth]{images/Kernels/EPI1.pdf}
        \end{minipage} &
        \begin{minipage}{1.0cm}
            \includegraphics[width=1.0\textwidth]{images/Kernels/EPI2.pdf}
        \end{minipage} &
        \begin{minipage}{1.0cm}
            \includegraphics[width=1.0\textwidth]{images/Kernels/EPI3.pdf}
        \end{minipage} &
        \begin{minipage}{1.0cm}
            \includegraphics[width=1.0\textwidth]{images/Kernels/Alpha.pdf}
        \end{minipage} &
        \begin{minipage}{1.0cm}
            \includegraphics[width=1.0\textwidth]{images/Kernels/Beta.pdf}
        \end{minipage} &
        \begin{minipage}{1.0cm}
            \includegraphics[width=1.0\textwidth]{images/Kernels/Gamma.pdf}
        \end{minipage} &
        \begin{minipage}{1.0cm}
            \includegraphics[width=1.0\textwidth]{images/Kernels/Dup1.pdf}
        \end{minipage} &
        \begin{minipage}{1.0cm}
            \includegraphics[width=1.0\textwidth]{images/Kernels/Dup2.pdf}
        \end{minipage} &
        \begin{minipage}{1.0cm}
            \includegraphics[width=1.0\textwidth]{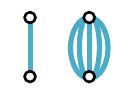}
        \end{minipage} &
        \begin{minipage}{1.0cm}
            \includegraphics[width=1.0\textwidth]{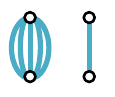}
        \end{minipage} 
        \\ \hline
        \#Connections      & 2                      & 2                       & 2                       & 4                        & 4                         & 4                         & 6                             & 4                             & 4                         & 6                         & 6                         \\ \hline
        PSNR                & 35.54                  & 35.51                   & 35.50                   & 35.85                    & 35.80                     & 35.79                     & \underline{36.07}             & 35.67                         & 35.76                     & 35.49                     & 35.89                     \\ \hline
        SSIM                & 0.9723                 & 0.9722                  & 0.9720                  & 0.9741                   & 0.9731                    & 0.9738                    & \underline{0.9752}            & 0.9728                        & 0.9730                    & 0.9723                    & 0.9743                    \\ \hline
        \end{tabular}
    }
\end{table*}
\begin{figure*}[ht]
    \centering
    \tabcolsep=0.03cm
    \renewcommand{\arraystretch}{0.4}
    \begin{tabu}{ccccccccccc}
        &&&
        LR &
        $\mathcal{K}^{SAS}$ &
        $\mathcal{K}^{EPI1}$ &
        $\mathcal{K}^{EPI2}$ &
        $\mathcal{K}^{Dup1-6}$ &
        $\mathcal{K}^{Dup2-6}$ &
        $\mathcal{K}^{\gamma}$ &
        \small{Ground-truth}
        \\

        \multirow{3}{*}{\rotatebox{90}{\hspace{43pt}\textit{general\_51}}} &
        \multirow{3}{*}[0.74cm]{\includegraphics[width=0.18\textwidth, height=2.145cm] {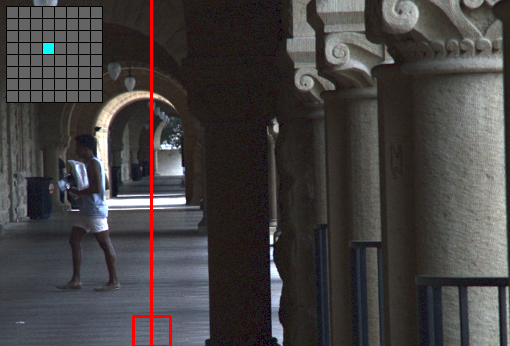}} &
        \hspace{0.001cm} &
        \includegraphics[width=0.090\textwidth, height=1.20cm] {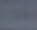} &
        \includegraphics[width=0.090\textwidth, height=1.20cm] {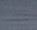} &
        \includegraphics[width=0.090\textwidth, height=1.20cm] {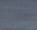} &
        \includegraphics[width=0.090\textwidth, height=1.20cm] {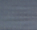} &
        \includegraphics[width=0.090\textwidth, height=1.20cm] {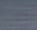} &
        \includegraphics[width=0.090\textwidth, height=1.20cm] {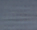} &
        \includegraphics[width=0.090\textwidth, height=1.20cm] {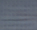} &
        \includegraphics[width=0.090\textwidth, height=1.20cm] {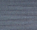} \\
        &&&
        \includegraphics[width=0.090\textwidth, height=0.42cm] {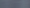} &
        \includegraphics[width=0.090\textwidth, height=0.42cm] {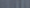} &
        \includegraphics[width=0.090\textwidth, height=0.42cm] {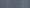} &
        \includegraphics[width=0.090\textwidth, height=0.42cm] {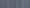} &
        \includegraphics[width=0.090\textwidth, height=0.42cm] {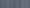} &
        \includegraphics[width=0.090\textwidth, height=0.42cm] {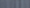} &
        \includegraphics[width=0.090\textwidth, height=0.42cm] {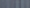} &
        \includegraphics[width=0.090\textwidth, height=0.42cm] {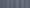} \\
        &&&
        \includegraphics[width=0.090\textwidth, height=0.42cm] {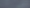} &
        \includegraphics[width=0.090\textwidth, height=0.42cm] {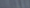} &
        \includegraphics[width=0.090\textwidth, height=0.42cm] {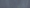} &
        \includegraphics[width=0.090\textwidth, height=0.42cm] {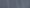} &
        \includegraphics[width=0.090\textwidth, height=0.42cm] {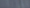} &
        \includegraphics[width=0.090\textwidth, height=0.42cm] {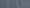} &
        \includegraphics[width=0.090\textwidth, height=0.42cm] {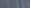} &
        \includegraphics[width=0.090\textwidth, height=0.42cm] {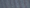} \\

        \vspace{-0.10cm} \\

        \multirow{3}{*}{\rotatebox{90}{\hspace{46pt}\textit{general\_46}}} &
        \multirow{3}{*}[0.74cm]{\includegraphics[width=0.18\textwidth, height=2.145cm] {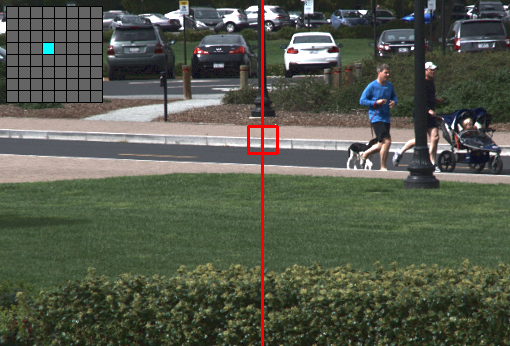}} &
        \hspace{0.001cm} &
        \includegraphics[width=0.090\textwidth, height=1.20cm] {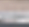} &
        \includegraphics[width=0.090\textwidth, height=1.20cm] {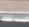} &
        \includegraphics[width=0.090\textwidth, height=1.20cm] {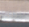} &
        \includegraphics[width=0.090\textwidth, height=1.20cm] {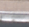} &
        \includegraphics[width=0.090\textwidth, height=1.20cm] {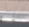} &
        \includegraphics[width=0.090\textwidth, height=1.20cm] {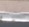} &
        \includegraphics[width=0.090\textwidth, height=1.20cm] {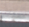} &
        \includegraphics[width=0.090\textwidth, height=1.20cm] {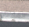} \\
        &&&
        \includegraphics[width=0.090\textwidth, height=0.42cm] {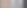} &
        \includegraphics[width=0.090\textwidth, height=0.42cm] {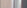} &
        \includegraphics[width=0.090\textwidth, height=0.42cm] {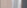} &
        \includegraphics[width=0.090\textwidth, height=0.42cm] {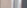} &
        \includegraphics[width=0.090\textwidth, height=0.42cm] {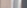} &
        \includegraphics[width=0.090\textwidth, height=0.42cm] {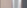} &
        \includegraphics[width=0.090\textwidth, height=0.42cm] {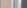} &
        \includegraphics[width=0.090\textwidth, height=0.42cm] {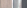} \\
        &&&
        \includegraphics[width=0.090\textwidth, height=0.42cm] {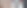} &
        \includegraphics[width=0.090\textwidth, height=0.42cm] {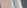} &
        \includegraphics[width=0.090\textwidth, height=0.42cm] {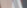} &
        \includegraphics[width=0.090\textwidth, height=0.42cm] {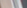} &
        \includegraphics[width=0.090\textwidth, height=0.42cm] {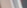} &
        \includegraphics[width=0.090\textwidth, height=0.42cm] {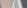} &
        \includegraphics[width=0.090\textwidth, height=0.42cm] {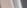} &
        \includegraphics[width=0.090\textwidth, height=0.42cm] {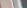} \\

        \vspace{-0.10cm} \\

        \multirow{3}{*}{\rotatebox{90}{\hspace{46pt}\textit{general\_56}}} &
        \multirow{3}{*}[0.74cm]{\includegraphics[width=0.18\textwidth, height=2.145cm] {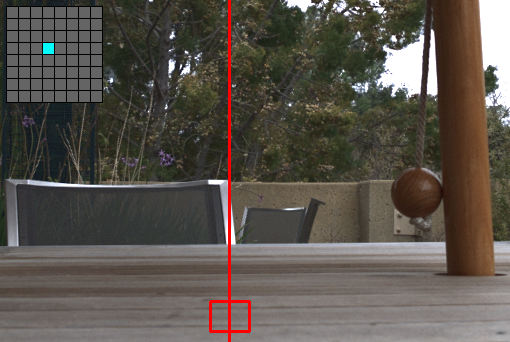}} &
        \hspace{0.001cm} &
        \includegraphics[width=0.090\textwidth, height=1.20cm] {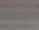} &
        \includegraphics[width=0.090\textwidth, height=1.20cm] {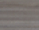} &
        \includegraphics[width=0.090\textwidth, height=1.20cm] {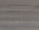} &
        \includegraphics[width=0.090\textwidth, height=1.20cm] {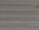} &
        \includegraphics[width=0.090\textwidth, height=1.20cm] {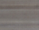} &
        \includegraphics[width=0.090\textwidth, height=1.20cm] {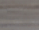} &
        \includegraphics[width=0.090\textwidth, height=1.20cm] {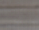} &
        \includegraphics[width=0.090\textwidth, height=1.20cm] {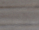} \\
        &&&
        \includegraphics[width=0.090\textwidth, height=0.42cm] {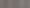} &
        \includegraphics[width=0.090\textwidth, height=0.42cm] {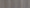} &
        \includegraphics[width=0.090\textwidth, height=0.42cm] {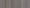} &
        \includegraphics[width=0.090\textwidth, height=0.42cm] {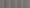} &
        \includegraphics[width=0.090\textwidth, height=0.42cm] {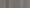} &
        \includegraphics[width=0.090\textwidth, height=0.42cm] {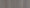} &
        \includegraphics[width=0.090\textwidth, height=0.42cm] {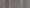} &
        \includegraphics[width=0.090\textwidth, height=0.42cm] {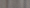} \\
        &&&
        \includegraphics[width=0.090\textwidth, height=0.42cm] {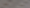} &
        \includegraphics[width=0.090\textwidth, height=0.42cm] {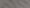} &
        \includegraphics[width=0.090\textwidth, height=0.42cm] {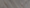} &
        \includegraphics[width=0.090\textwidth, height=0.42cm] {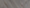} &
        \includegraphics[width=0.090\textwidth, height=0.42cm] {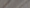} &
        \includegraphics[width=0.090\textwidth, height=0.42cm] {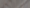} &
        \includegraphics[width=0.090\textwidth, height=0.42cm] {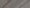} &
        \includegraphics[width=0.090\textwidth, height=0.42cm] {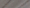} \\

        \vspace{-0.10cm} \\

        \multirow{3}{*}{\rotatebox{90}{\hspace{45pt}\textit{general\_9}}} &
        \multirow{3}{*}[0.74cm]{\includegraphics[width=0.18\textwidth, height=2.145cm] {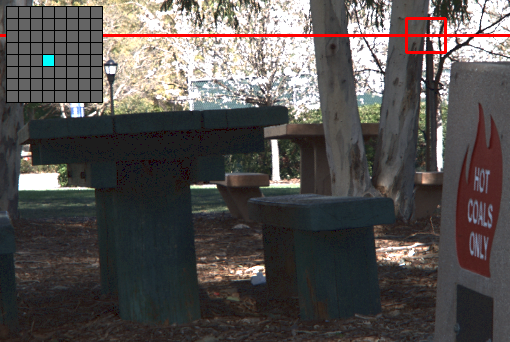}} &
        \hspace{0.001cm} &
        \includegraphics[width=0.090\textwidth, height=1.20cm] {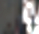} &
        \includegraphics[width=0.090\textwidth, height=1.20cm] {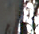} &
        \includegraphics[width=0.090\textwidth, height=1.20cm] {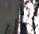} &
        \includegraphics[width=0.090\textwidth, height=1.20cm] {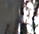} &
        \includegraphics[width=0.090\textwidth, height=1.20cm] {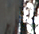} &
        \includegraphics[width=0.090\textwidth, height=1.20cm] {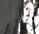} &
        \includegraphics[width=0.090\textwidth, height=1.20cm] {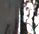} &
        \includegraphics[width=0.090\textwidth, height=1.20cm] {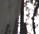} \\
        &&&
        \includegraphics[width=0.090\textwidth, height=0.42cm] {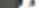} &
        \includegraphics[width=0.090\textwidth, height=0.42cm] {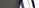} &
        \includegraphics[width=0.090\textwidth, height=0.42cm] {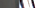} &
        \includegraphics[width=0.090\textwidth, height=0.42cm] {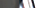} &
        \includegraphics[width=0.090\textwidth, height=0.42cm] {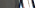} &
        \includegraphics[width=0.090\textwidth, height=0.42cm] {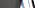} &
        \includegraphics[width=0.090\textwidth, height=0.42cm] {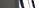} &
        \includegraphics[width=0.090\textwidth, height=0.42cm] {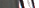} \\
        &&&
        \includegraphics[width=0.090\textwidth, height=0.42cm] {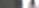} &
        \includegraphics[width=0.090\textwidth, height=0.42cm] {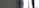} &
        \includegraphics[width=0.090\textwidth, height=0.42cm] {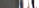} &
        \includegraphics[width=0.090\textwidth, height=0.42cm] {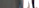} &
        \includegraphics[width=0.090\textwidth, height=0.42cm] {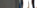} &
        \includegraphics[width=0.090\textwidth, height=0.42cm] {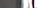} &
        \includegraphics[width=0.090\textwidth, height=0.42cm] {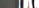} &
        \includegraphics[width=0.090\textwidth, height=0.42cm] {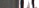} \\

    \end{tabu}
    \caption{Visualization of decomposition kernels. Two EPIs are extracted along the red line and the two angular dimensions in the red box separately.}
    \label{fig:QualKernels}
\end{figure*}

\subsubsection{Non-down-sampled Images}
While most LFSR methods are performing on down-sampled images, \textit{i.e.} recovering the resolution, the training process might lead the model to overfit the down-sampling method. One way to verify this is to test the models on the natural images without down-sampling. Similar to the previous qualitative comparison, we perform DKNet and Yeung \textit{et al.} \cite{yeungSAS_LFSR2019} on the original LF images, and a few of them are visualized in Fig. \ref{fig:FurtherEnhancement}.

Looking at the zoom-in SAIs, some detailed differences can be observed such as the edges of leaves in \textit{occlusion\_9} and branches in \textit{occlusion\_42}, which are sharper in DKNet than the other super-solved image. In \textit{reflective\_18}, the block edges and the shadow on the arch are visibly enhanced by DKNet too compared to Yeung \textit{et al.} (4D) \cite{yeungSAS_LFSR2019} and the original image.

The visualized EPIs provide more insight behind the surface. While Yeung \textit{et al.} (4D) \cite{yeungSAS_LFSR2019} barely makes any change, the EPI patterns are enhanced by DKNet with sharp details in \textit{occlusion\_42}. The EPIs of Yeung \textit{et al.} (4D) \cite{yeungSAS_LFSR2019} in \textit{occlusion\_9} and \textit{reflective\_18} even reveal the destroyed LF structure as the EPI patterns are changed while DKNet well preserves the LF structure without making structural changes to the EPIs.

These findings suggest the potential generalization problem of Yeung \textit{et al.} (4D) \cite{yeungSAS_LFSR2019} despite its high performance when recovering down-sampled LF images. By contrast, with the strength of better generalization, DKNet can yield excellent performance in not only recovering down-sampled images but also better super-resolving non-down-sampled ones.

\subsection{Ablation Studies} \label{section:Ablation}

\subsubsection{Decomposition Kernels Types}
To discover the impact of different decomposition kernels, we train DKNet with the proposed kernels separately in scale $4 \times$, while the other configurations remain the same. Miscellaneous dataset \cite{StanfordLytro} is used as the test dataset hereafter. Regarding the duplication kernels, we train two variants for both types, namely the 4-connection $\mathcal{K}^{Dup1-4}$ and $\mathcal{K}^{Dup2-4}$ with 3 duplicated $(x, y)$ or $(u, v)$ connections, and likewise, the 6-connection $\mathcal{K}^{Dup1-6}$ and $\mathcal{K}^{Dup2-6}$ with 5 duplicated connections. The quantitative results are demonstrated in Table \ref{tab:PerfKernel}. 

A general trend is observed that except the duplication kernels, more connections lead to better performance. While the 2-connection kernels $\mathcal{K}^{SAS}$, $\mathcal{K}^{EPI1}$ and $\mathcal{K}^{EPI2}$ achieve 35.50 dB to 35.54 dB PSNR, the 4-connection variants combining any two of them, namely $\mathcal{K}^{\alpha}$, $\mathcal{K}^{\beta}$ and $\mathcal{K}^{EPI3}$, can obtain about 0.30 dB extra, resulting in 35.79 dB to 35.85 dB. This benefit brought by increasing and diversifying connections can be further extended when all possible connections are exploited in $\mathcal{K}^{\gamma}$, which achieves the best result of 36.07 dB.

It is worth noticing that the EPI-based methods \cite{wuEPICNN_TPAMI2018} are usually inferior to the SAS-based method \cite{yeungSAS_ECCV2018} in LF reconstruction in terms of performance. However, DKNet embedded with the EPI-based kernels can achieve similar performance with $\mathcal{K}^{SAS}$. We believe there are two important implications behind this result. The first one is DKNet's compatibility to take advantage of both EPI-based and SAS-based decomposition kernels in a unified manner. The second one is that the spatial, angular, and EPI sub-spaces are of equal importance for LFSR, as well as inter- and intra-domain connections.

On the other hand, we find that the performance gain brought by increasing connections does not persist among the duplication kernels, which produce relatively worse results when compared to their counterparts with the same number of connections. With 4 connections, $\mathcal{K}^{Dup1-4}$ and $\mathcal{K}^{Dup2-4}$ achieve 35.67 dB and 35.76 dB respectively, which are slightly lower when compared to $\mathcal{K}^{EPI3}$, $\mathcal{K}^{\alpha}$ and $\mathcal{K}^{\beta}$. With 6 connections, $\mathcal{K}^{Dup1-6}$ achieves 35.49 dB, worse than its 4-connection version $\mathcal{K}^{Dup1-4}$ by 0.18 dB, which we attribute to its over-reliance on a specific sub-space. The 6-connection $\mathcal{K}^{Dup2-6}$ achieves 35.89 dB, which is better than its 4-connection version $\mathcal{K}^{Dup2-4}$ by 0.13 dB, but still falls behind its 6-connection counterpart, $\mathcal{K}^{\gamma}$, by 0.18 dB. The above results suggest that besides the number, the diversity of connections actually is more important for comprehensive spatio-angular feature extraction.

We depict some visual results in Fig. \ref{fig:QualKernels} to delve into the impact of specific decomposition kernels. In \textit{general\_51}, $\mathcal{K}^{EPI1}$ cannot recover the floor patterns but generate a textureless area while the other kernels recover this area to some extent. It could be seen in its EPIs as well that the patterns are missing. Similarly, in \textit{general\_46}, the kernels generally have recovered the edge of the road except $\mathcal{K}^{EPI1}$ and $\mathcal{K}^{Dup2-6}$ produce blurry results in the red box. Their second EPIs also present some hallucinatory patterns. The failure of $\mathcal{K}^{EPI1}$ and $\mathcal{K}^{Dup2-6}$ implies their deficient spatio-angular representation, while $\mathcal{K}^{EPI1}$ is caused by its inadequate connections to exploit the information of other sub-spaces, and $\mathcal{K}^{Dup2-6}$ is due to its unbalanced connections.

When looking at \textit{general\_56}, some wood grains in the red box are spotted that cannot be captured by any 2-connection kernels. Looking at the SAIs, $\mathcal{K}^{SAS}$, $\mathcal{K}^{EPI1}$ and $\mathcal{K}^{EPI2}$ generate incorrect artifacts in this part. On the contrary, $\mathcal{K}^{\gamma}$, $\mathcal{K}^{Dup1-6}$ and $\mathcal{K}^{Dup2-6}$ are able to obtain the correct texture. However, the EPIs portray more insight from another perspective. While the 2-connection kernels just present incorrect edges in the first EPIs, the patterns in the second EPIs are completely in the wrong direction when compared with the ground-truth. The duplication kernels $\mathcal{K}^{Dup2-6}$ seems to generate roughly correct structure if we just look at its single SAI, but hallucination is spotted in its second EPI, which means the structure of the reconstructed LF image is flawed. Therefore, $\mathcal{K}^{Dup1-6}$ and $\mathcal{K}^{\gamma}$ are the only kernels that can correctly super-resolve this sample.

The improvement brought by $\mathcal{K}^{\gamma}$ is more significant in \textit{general\_9} where occlusion occurs between the two trees in the red box. All the kernels cannot generate distinguishable edges between the tree trunks except $\mathcal{K}^{\gamma}$. It is also observed that only $\mathcal{K}^{\gamma}$ can produce sharp edges in both EPIs, whereas the others portray abnormal patterns.

\begin{figure}[t!]
    \centering
    \tabcolsep=0.0cm
    \begin{tabu}{cc}
        \includegraphics[width=0.246\textwidth]{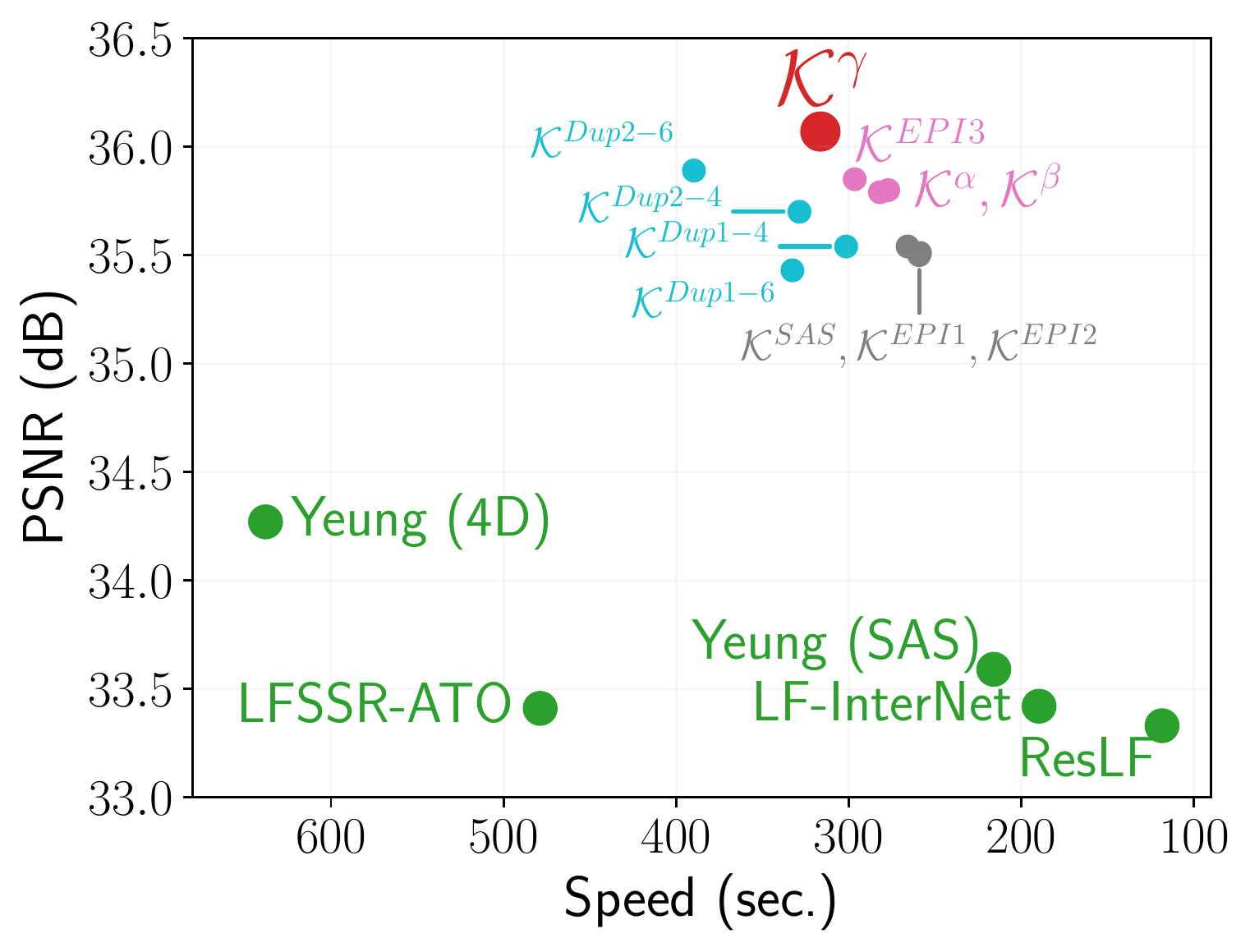} &
        \includegraphics[width=0.246\textwidth]{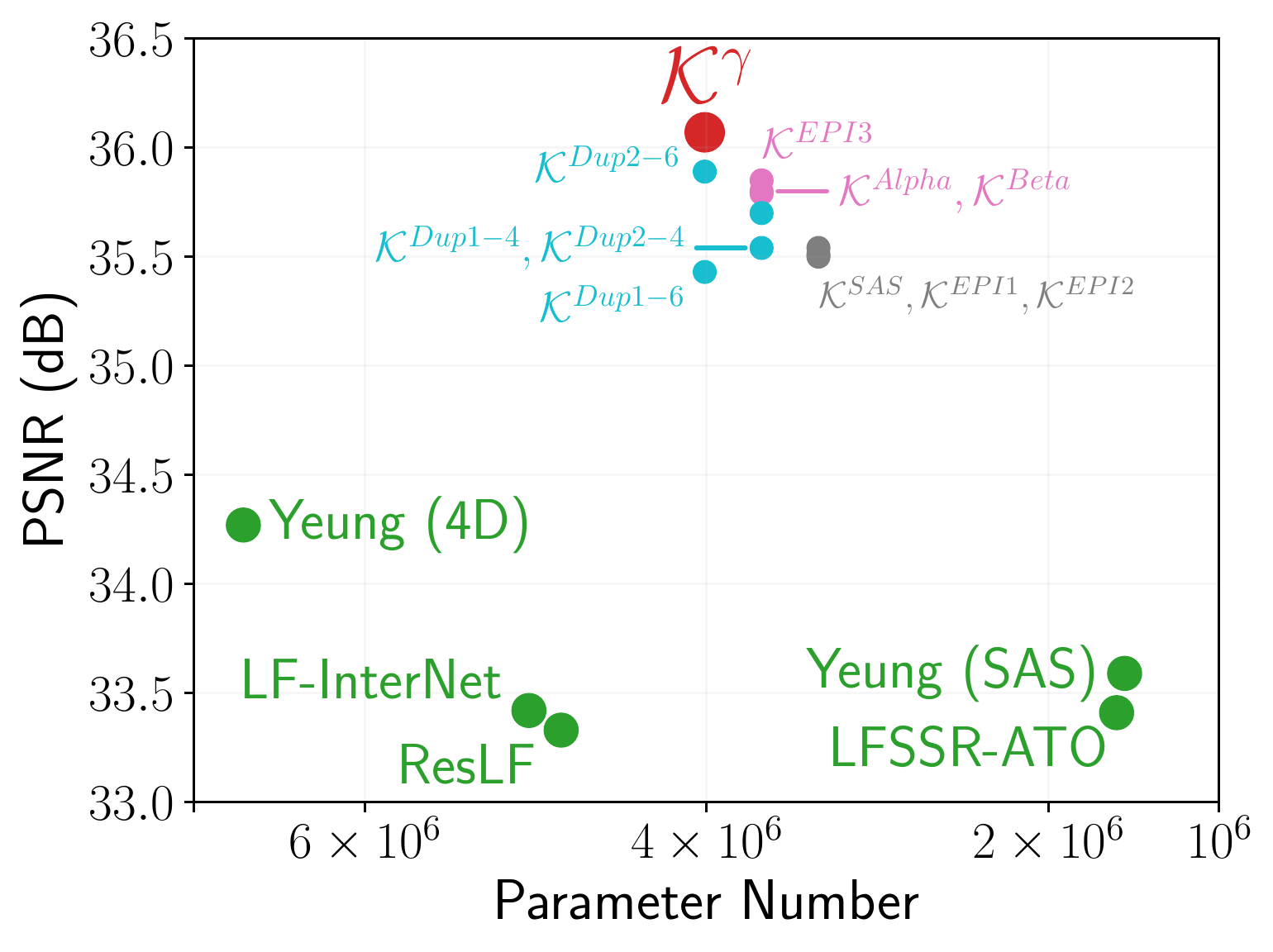} \\
        \multicolumn{2}{c}{(a) Decomposition Kernel Types $(L=18)$.} \\

        \includegraphics[width=0.246\textwidth]{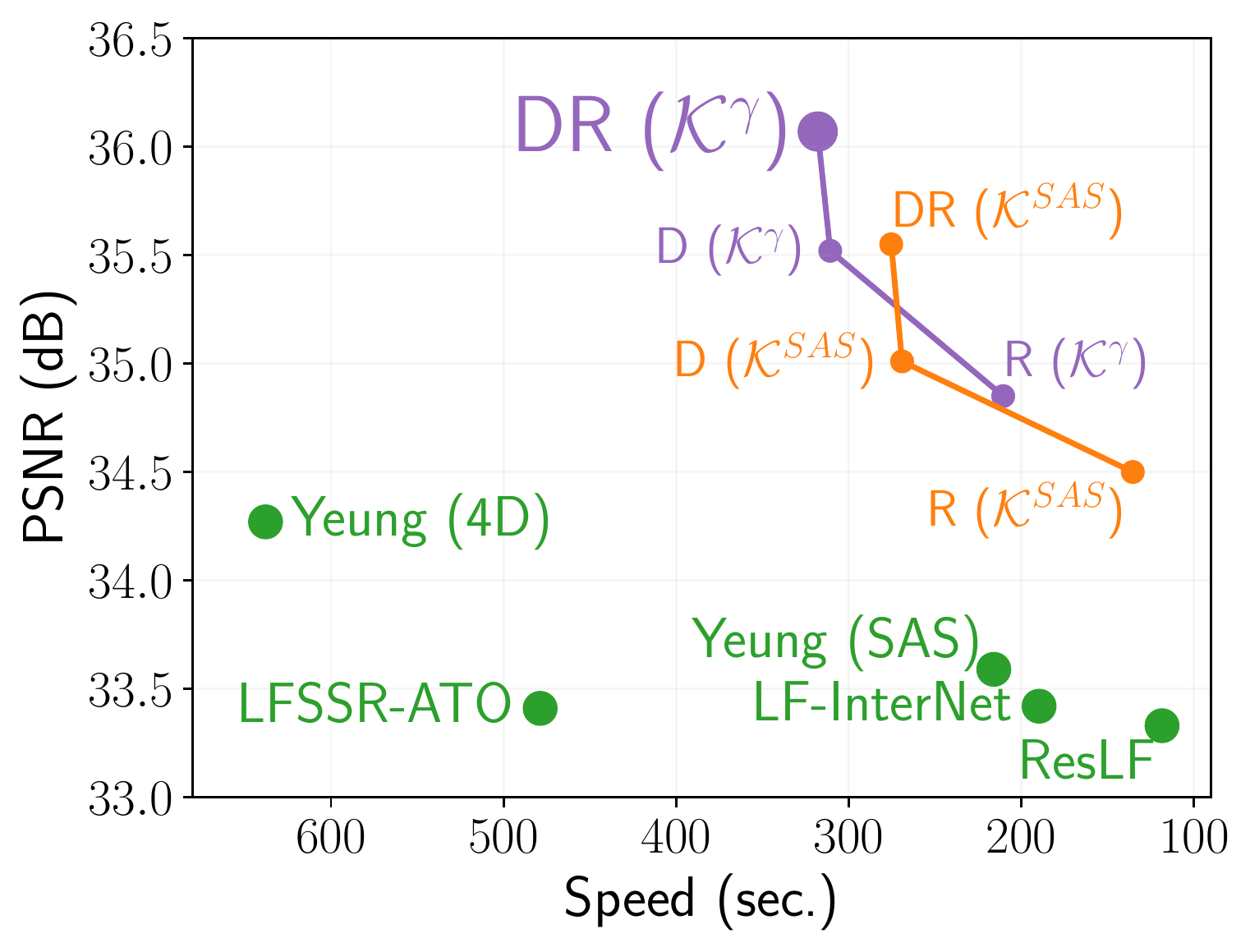} &
        \includegraphics[width=0.246\textwidth]{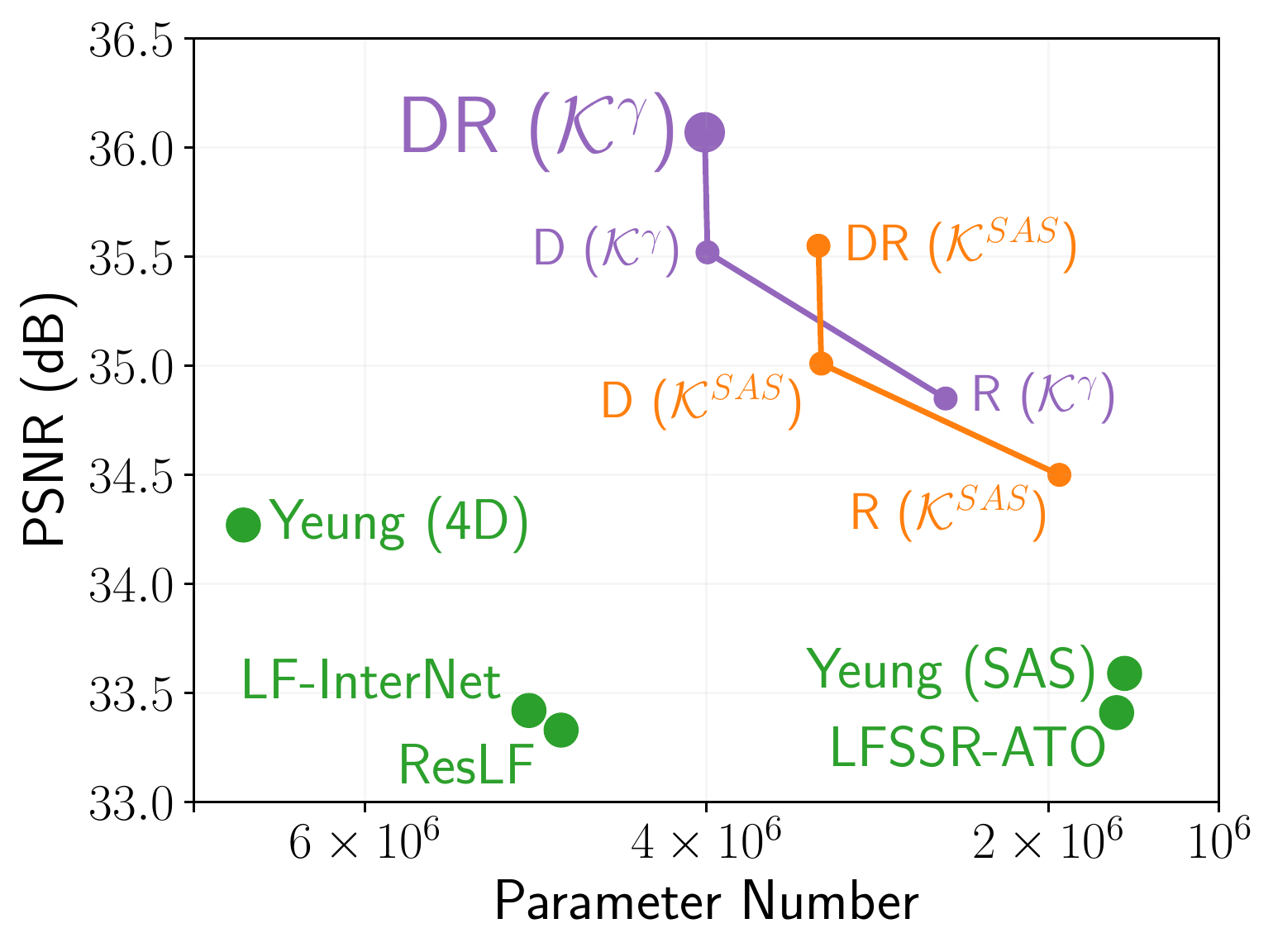} \\
        \multicolumn{2}{c}{(b) Dense and raw image connections $(L=18)$.} \\

        \includegraphics[width=0.246\textwidth]{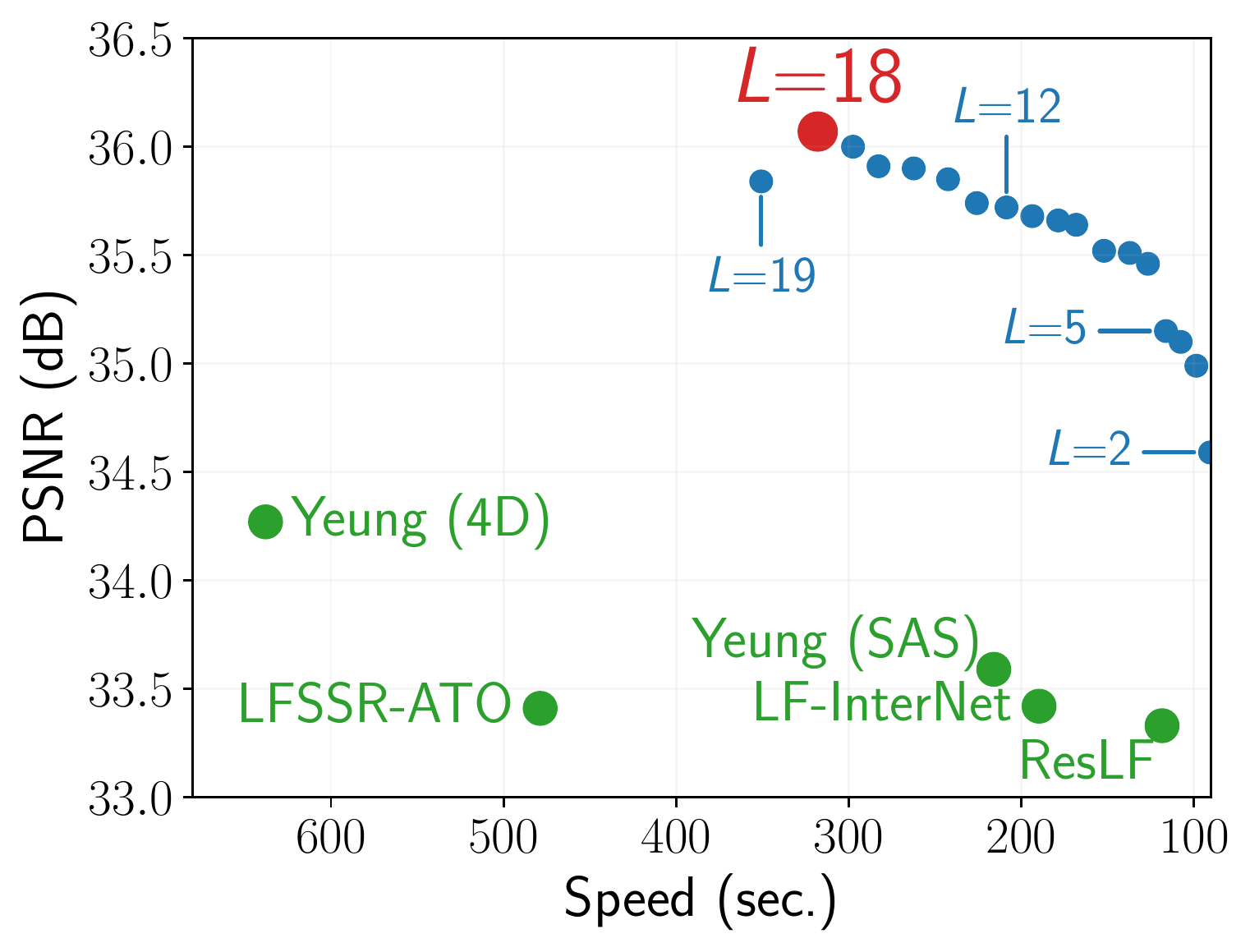} &
        \includegraphics[width=0.246\textwidth]{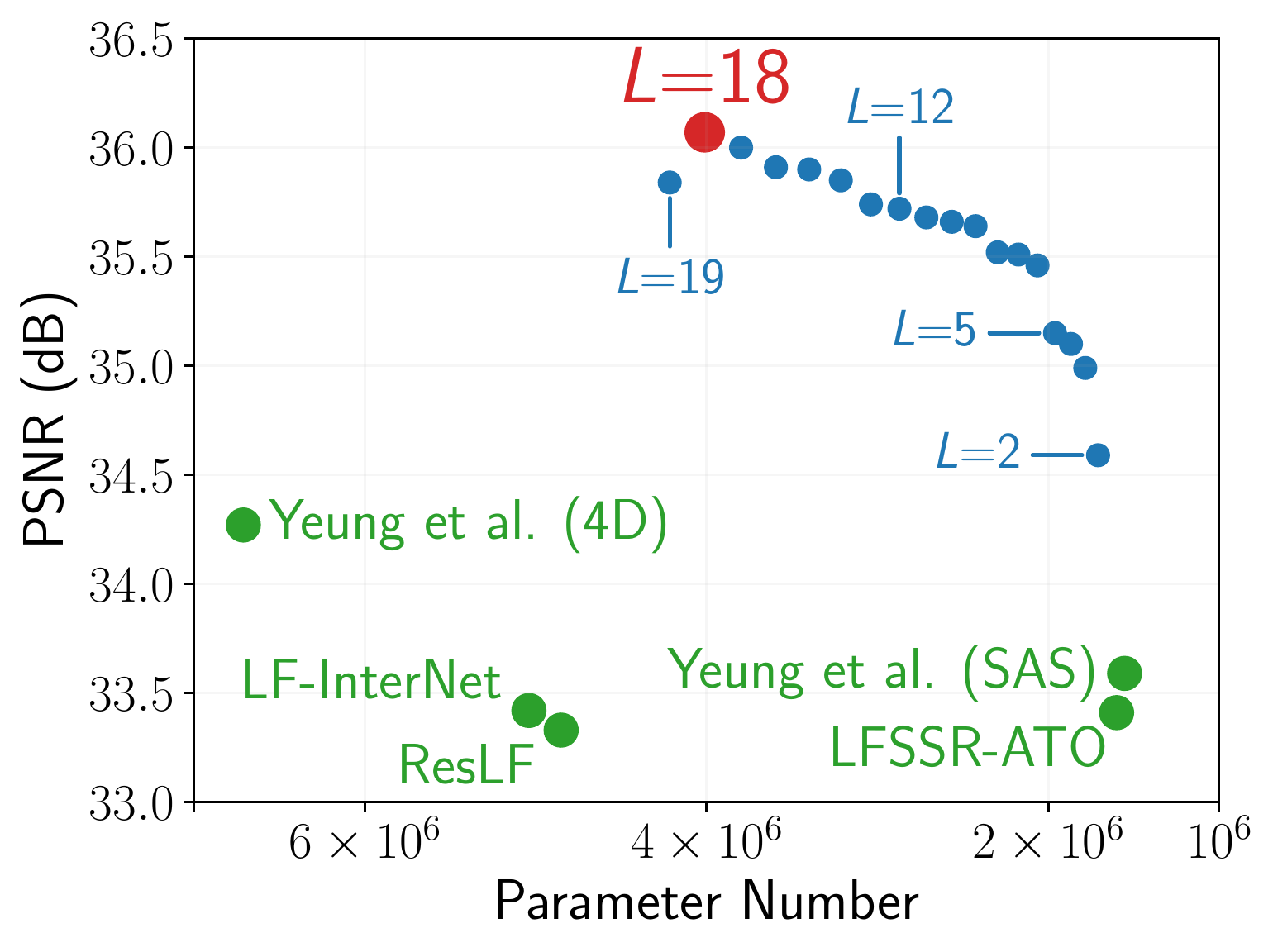} \\
        \multicolumn{2}{c}{(c) Decomposition kernel numbers $(\mathcal{K}^{\gamma})$.} \\

        \includegraphics[width=0.246\textwidth]{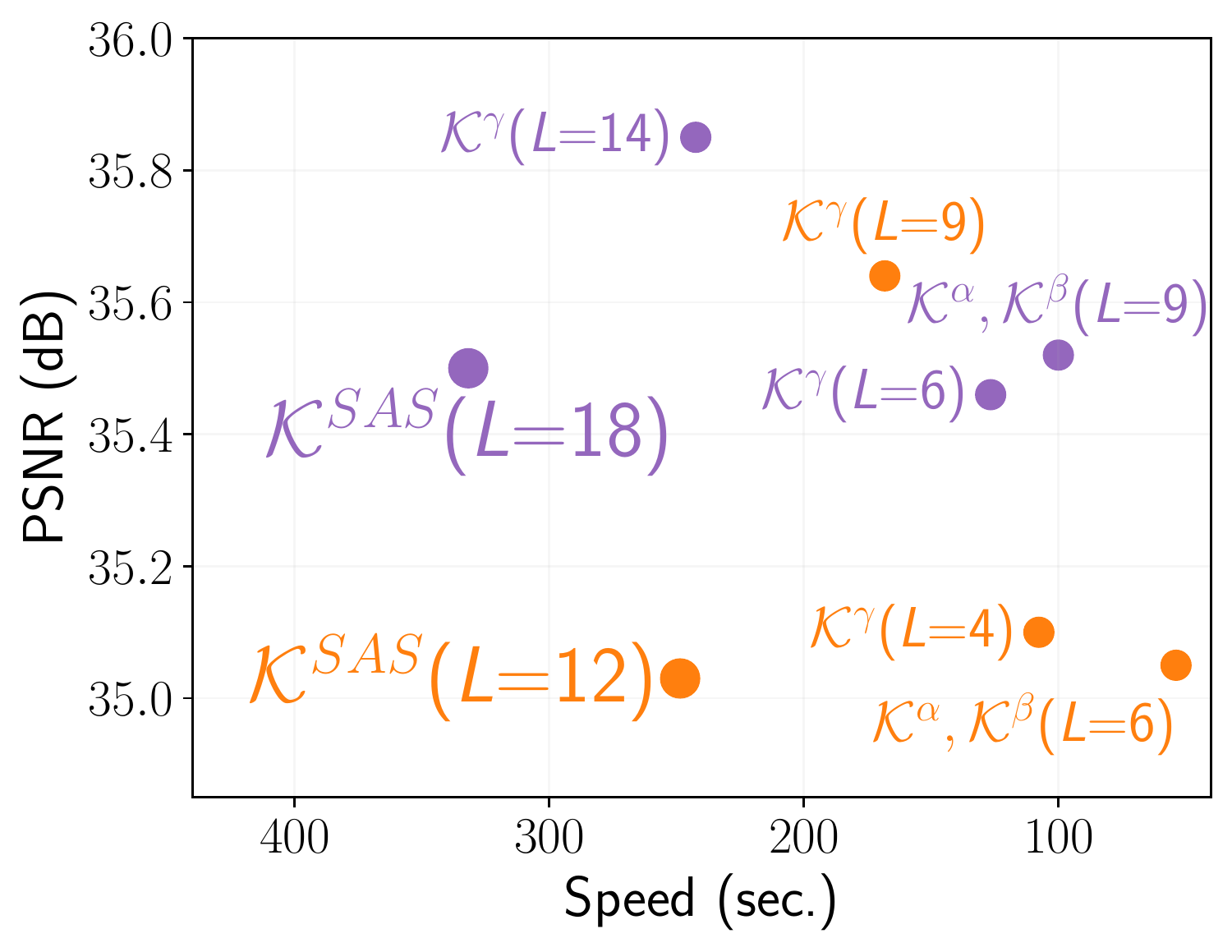} &
        \includegraphics[width=0.246\textwidth]{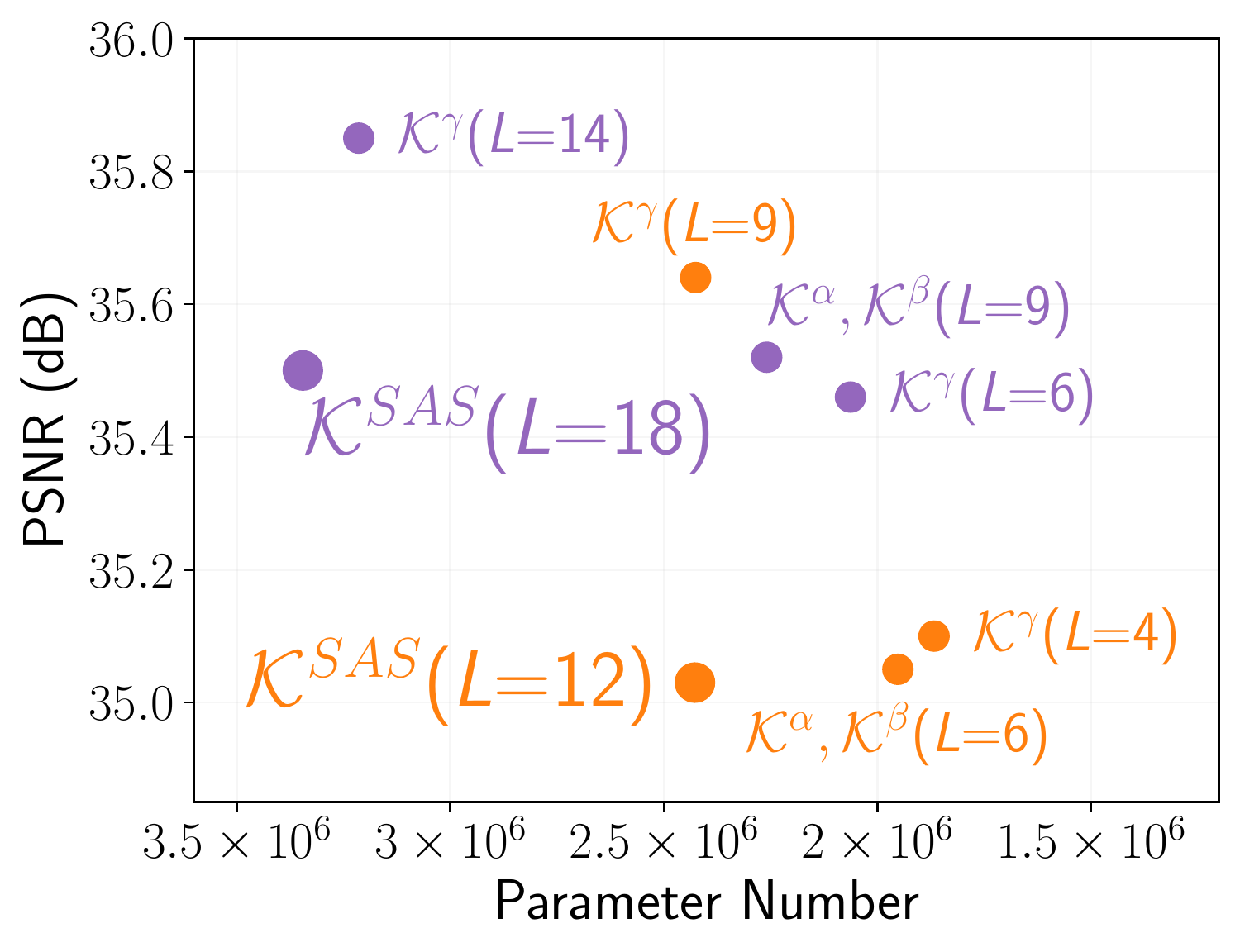} \\
        \multicolumn{2}{c}{(d) Decomposition kernel types and numbers.} \\
    \end{tabu}
    \caption{Plots of performance versus speed and trainable parameter numbers under different conditions. Points closer to the right signify higher speed or fewer parameters while the upper points mean better performance. Therefore, points closer to the top-right corners indicate better efficiency.}
    \vspace{-0.3cm}
    \label{fig:PerfNCost}
\end{figure}

The accurate reconstruction of $\mathcal{K}^{\gamma}$ once again confirms the importance of connection diversity for LFSR to capture visual patterns in a wide range of sub-spaces simultaneously. Moreover, although the generated SAIs may seem good, their EPIs, especially the neglected EPI sub-spaces, can reveal the problems hidden beneath the LF structure. They can be hallucination by $\mathcal{K}^{EPI1}$ and $\mathcal{K}^{Dup2-6}$ in \textit{general\_46}, and by most kernels in \textit{general\_9}. They can be the entirely missing patterns by $\mathcal{K}^{EPI1}$ in \textit{general\_51}. Or, they can be completely misdirected patterns by the 2-connection kernels in \textit{general\_56}. These differences suggest the potential application by using EPIs of extended sub-spaces for LFSR error diagnosis in future research.

Finally, we investigate the efficiency of the kernels. Two plots are illustrated in Fig. \ref{fig:PerfNCost} (a) to demonstrate the performance's trade-off with the speed and parameter numbers correspondingly. All the methods in the previous experiment are included except HDDRNet \cite{mengHDDRNet_AAAI2020} as its PSNR result is too low. Generally, all the kernels fall between Yeung \textit{et al.}'s 4D and SAS versions in terms of the costs of runtime (speed) and memory (parameter number) while outperforming the best 4D version of Yeung \textit{et al.} by at least 1.22 dB. In consideration of the balance of performance and cost, there are three primary groups for decomposition kernel selection. The first group consists of the three 2-connection kernels, namely $\mathcal{K}^{SAS}$, $\mathcal{K}^{EPI1}$ and $\mathcal{K}^{EPI2}$, as their running speed is close to Yeung \textit{et al.}'s SAS version and have a relatively smaller model size while obtaining a considerable performance in terms of PSNR. The second group is $\mathcal{K}^{\gamma}$, which achieves the best PSNR. And the third group is the trade-off solution that consists of the 4-connection combination kernels, namely $\mathcal{K}^{EPI3}$, $\mathcal{K}^{\alpha}$ and $\mathcal{K}^{\beta}$. Because of the deficiency in connection diversity, the duplication kernels are at a disadvantage as they perform worse than their counterparts given the same cost.

\begin{figure*}[t!]
    \centering
    \resizebox{\textwidth}{!}{
    \tabcolsep=0.02cm
    \renewcommand{\arraystretch}{0.4}
    \begin{tabular}{cccccccccccc}
        \multicolumn{1}{l}{}
        & \multicolumn{2}{c}{LR}
        & \multicolumn{2}{c}{SRGAN \cite{ledigSRGAN_CVPR2017}}
        & \multicolumn{2}{c}{DKNet}
        & \multicolumn{2}{c}{TE-DKNet}
        & \multicolumn{2}{c}{Ground-truth}
        \\
        
        \multicolumn{1}{c}{\rotatebox{90}{\hspace{12pt}{\textit{general\_17}}}}
        & \multicolumn{2}{l}{\includegraphics[width=0.19\textwidth, height=0.130\textwidth]{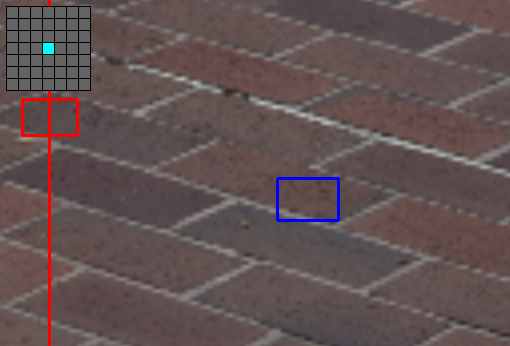}}
        & \multicolumn{2}{l}{\includegraphics[width=0.19\textwidth, height=0.130\textwidth]{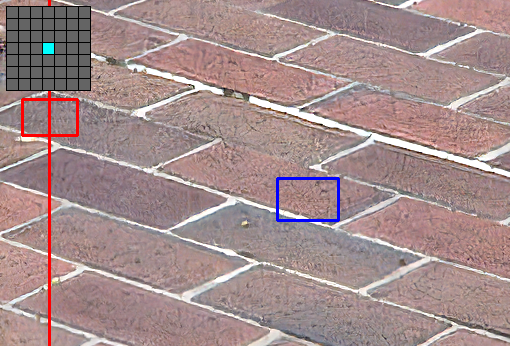}}
        & \multicolumn{2}{l}{\includegraphics[width=0.19\textwidth, height=0.130\textwidth]{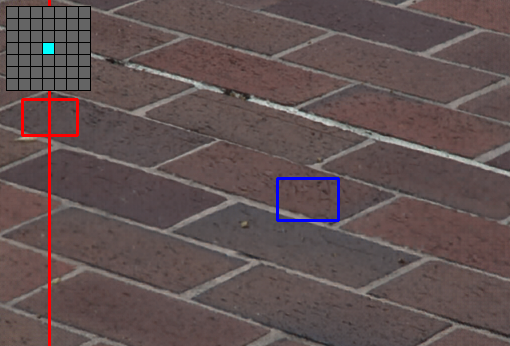}}
        & \multicolumn{2}{l}{\includegraphics[width=0.19\textwidth, height=0.130\textwidth]{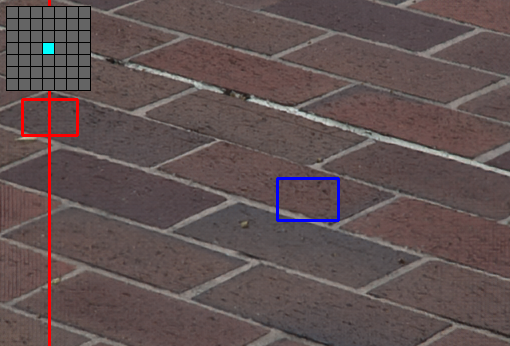}}
        & \multicolumn{2}{l}{\includegraphics[width=0.19\textwidth, height=0.130\textwidth]{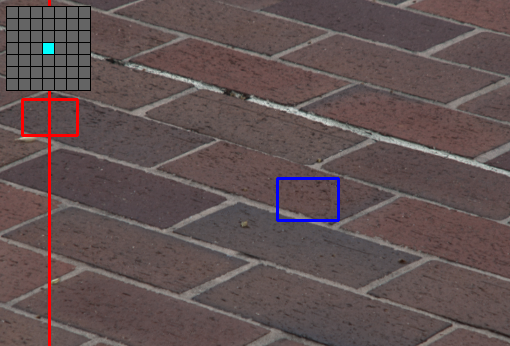}}
        &
        \\
        
        & \multicolumn{1}{l}{\includegraphics[width=0.09\textwidth, height=0.052\textwidth,cfbox=red 1pt 0pt] {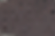}}
        & \multicolumn{1}{r}{\includegraphics[width=0.09\textwidth, height=0.052\textwidth,cfbox=blue 1pt 0pt]{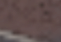}}
        & \multicolumn{1}{l}{\includegraphics[width=0.09\textwidth, height=0.052\textwidth,cfbox=red 1pt 0pt] {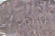}}
        & \multicolumn{1}{r}{\includegraphics[width=0.09\textwidth, height=0.052\textwidth,cfbox=blue 1pt 0pt]{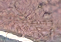}}
        & \multicolumn{1}{l}{\includegraphics[width=0.09\textwidth, height=0.052\textwidth,cfbox=red 1pt 0pt] {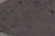}}
        & \multicolumn{1}{r}{\includegraphics[width=0.09\textwidth, height=0.052\textwidth,cfbox=blue 1pt 0pt]{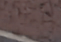}}
        & \multicolumn{1}{l}{\includegraphics[width=0.09\textwidth, height=0.052\textwidth,cfbox=red 1pt 0pt] {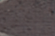}}
        & \multicolumn{1}{r}{\includegraphics[width=0.09\textwidth, height=0.052\textwidth,cfbox=blue 1pt 0pt]{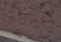}}
        & \multicolumn{1}{l}{\includegraphics[width=0.09\textwidth, height=0.052\textwidth,cfbox=red 1pt 0pt] {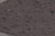}}
        & \multicolumn{1}{r}{\includegraphics[width=0.09\textwidth, height=0.052\textwidth,cfbox=blue 1pt 0pt]{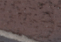}}
        &
        \\

        \multicolumn{1}{c}{}
        & \multicolumn{2}{l}{\includegraphics[width=0.19\textwidth, height=0.016\textheight]{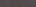}}
        & \multicolumn{2}{l}{\includegraphics[width=0.19\textwidth, height=0.016\textheight]{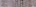}}
        & \multicolumn{2}{l}{\includegraphics[width=0.19\textwidth, height=0.016\textheight]{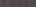}}
        & \multicolumn{2}{l}{\includegraphics[width=0.19\textwidth, height=0.016\textheight]{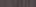}}
        & \multicolumn{2}{l}{\includegraphics[width=0.19\textwidth, height=0.016\textheight]{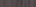}}
        &
        \\

        \multicolumn{1}{c}{}
        & \multicolumn{2}{l}{\includegraphics[width=0.19\textwidth, height=0.016\textheight]{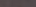}}
        & \multicolumn{2}{l}{\includegraphics[width=0.19\textwidth, height=0.016\textheight]{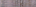}}
        & \multicolumn{2}{l}{\includegraphics[width=0.19\textwidth, height=0.016\textheight]{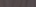}}
        & \multicolumn{2}{l}{\includegraphics[width=0.19\textwidth, height=0.016\textheight]{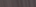}}
        & \multicolumn{2}{l}{\includegraphics[width=0.19\textwidth, height=0.016\textheight]{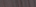}}
        &
        \\

        \multicolumn{1}{c}{\vspace{0.15cm}}
        & \multicolumn{2}{c}{32.19/0.8226}
        & \multicolumn{2}{c}{13.63/0.6259}
        & \multicolumn{2}{c}{38.70/0.9795}
        & \multicolumn{2}{c}{38.17/0.9773}
        & \multicolumn{2}{c}{}
        \\

        \multicolumn{1}{c}{\rotatebox{90}{\hspace{12pt}{\textit{general\_45}}}}
        & \multicolumn{2}{l}{\includegraphics[width=0.19\textwidth, height=0.130\textwidth]{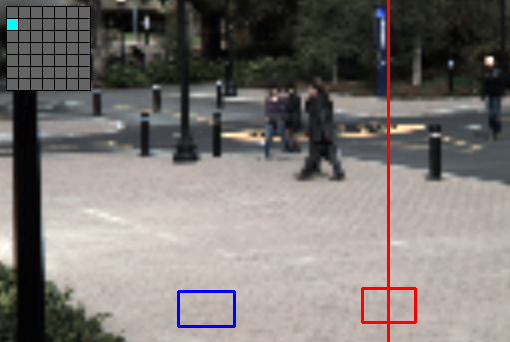}}
        & \multicolumn{2}{l}{\includegraphics[width=0.19\textwidth, height=0.130\textwidth]{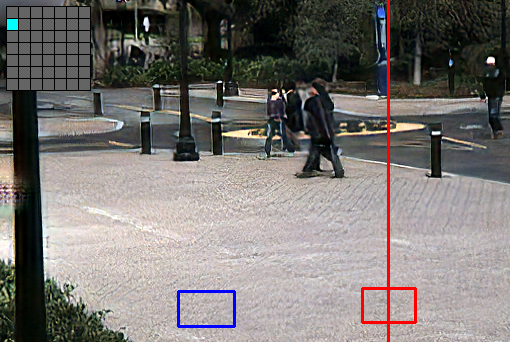}}
        & \multicolumn{2}{l}{\includegraphics[width=0.19\textwidth, height=0.130\textwidth]{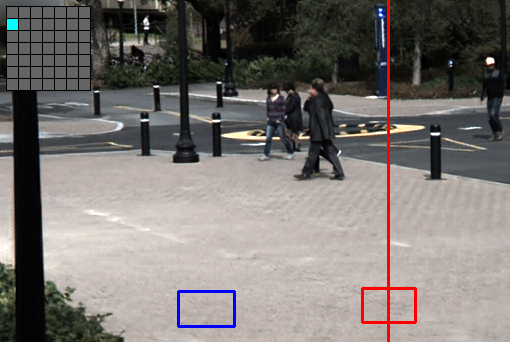}}
        & \multicolumn{2}{l}{\includegraphics[width=0.19\textwidth, height=0.130\textwidth]{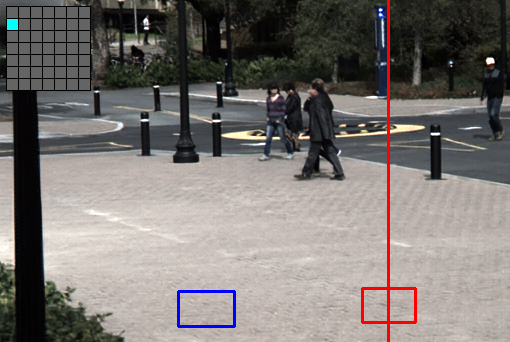}}
        & \multicolumn{2}{l}{\includegraphics[width=0.19\textwidth, height=0.130\textwidth]{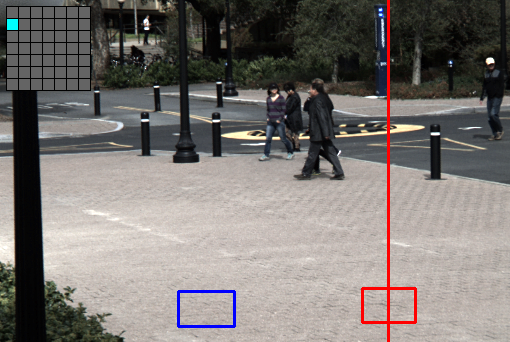}}
        &
        \\
        
        & \multicolumn{1}{l}{\includegraphics[width=0.09\textwidth, height=0.052\textwidth,cfbox=red 1pt 0pt] {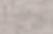}}
        & \multicolumn{1}{r}{\includegraphics[width=0.09\textwidth, height=0.052\textwidth,cfbox=blue 1pt 0pt]{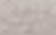}}
        & \multicolumn{1}{l}{\includegraphics[width=0.09\textwidth, height=0.052\textwidth,cfbox=red 1pt 0pt] {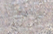}}
        & \multicolumn{1}{r}{\includegraphics[width=0.09\textwidth, height=0.052\textwidth,cfbox=blue 1pt 0pt]{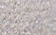}}
        & \multicolumn{1}{l}{\includegraphics[width=0.09\textwidth, height=0.052\textwidth,cfbox=red 1pt 0pt] {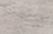}}
        & \multicolumn{1}{r}{\includegraphics[width=0.09\textwidth, height=0.052\textwidth,cfbox=blue 1pt 0pt]{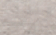}}
        & \multicolumn{1}{l}{\includegraphics[width=0.09\textwidth, height=0.052\textwidth,cfbox=red 1pt 0pt] {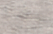}}
        & \multicolumn{1}{r}{\includegraphics[width=0.09\textwidth, height=0.052\textwidth,cfbox=blue 1pt 0pt]{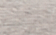}}
        & \multicolumn{1}{l}{\includegraphics[width=0.09\textwidth, height=0.052\textwidth,cfbox=red 1pt 0pt] {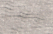}}
        & \multicolumn{1}{r}{\includegraphics[width=0.09\textwidth, height=0.052\textwidth,cfbox=blue 1pt 0pt]{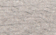}}
        &
        \\

        \multicolumn{1}{c}{}
        & \multicolumn{2}{l}{\includegraphics[width=0.19\textwidth, height=0.016\textheight]{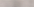}}
        & \multicolumn{2}{l}{\includegraphics[width=0.19\textwidth, height=0.016\textheight]{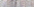}}
        & \multicolumn{2}{l}{\includegraphics[width=0.19\textwidth, height=0.016\textheight]{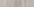}}
        & \multicolumn{2}{l}{\includegraphics[width=0.19\textwidth, height=0.016\textheight]{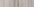}}
        & \multicolumn{2}{l}{\includegraphics[width=0.19\textwidth, height=0.016\textheight]{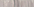}}
        &
        \\

        \multicolumn{1}{c}{}
        & \multicolumn{2}{l}{\includegraphics[width=0.19\textwidth, height=0.016\textheight]{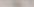}}
        & \multicolumn{2}{l}{\includegraphics[width=0.19\textwidth, height=0.016\textheight]{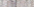}}
        & \multicolumn{2}{l}{\includegraphics[width=0.19\textwidth, height=0.016\textheight]{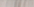}}
        & \multicolumn{2}{l}{\includegraphics[width=0.19\textwidth, height=0.016\textheight]{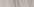}}
        & \multicolumn{2}{l}{\includegraphics[width=0.19\textwidth, height=0.016\textheight]{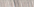}}
        &
        \\

        \multicolumn{1}{c}{\vspace{0.15cm}}
        & \multicolumn{2}{c}{30.19/0.8097}
        & \multicolumn{2}{c}{26.16/0.7886}
        & \multicolumn{2}{c}{37.32/0.9753}
        & \multicolumn{2}{c}{37.17/0.9749}
        & \multicolumn{2}{c}{}
        \\

        \multicolumn{1}{c}{\rotatebox{90}{\hspace{12pt}{\textit{general\_48}}}}
        & \multicolumn{2}{l}{\includegraphics[width=0.19\textwidth, height=0.130\textwidth]{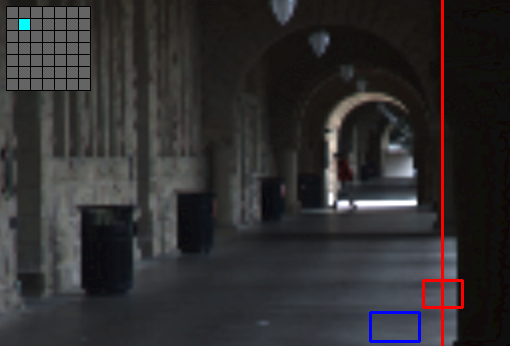}}
        & \multicolumn{2}{l}{\includegraphics[width=0.19\textwidth, height=0.130\textwidth]{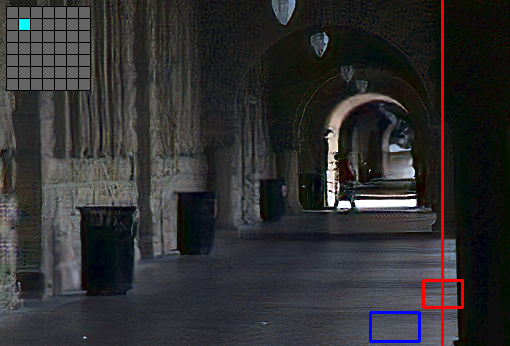}}
        & \multicolumn{2}{l}{\includegraphics[width=0.19\textwidth, height=0.130\textwidth]{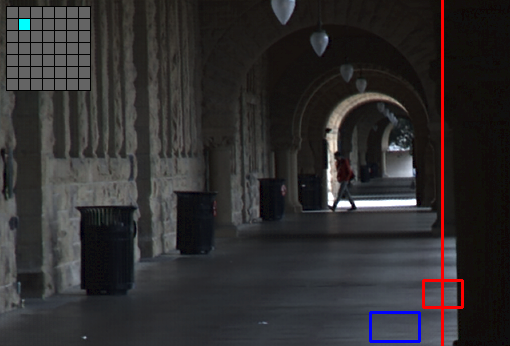}}
        & \multicolumn{2}{l}{\includegraphics[width=0.19\textwidth, height=0.130\textwidth]{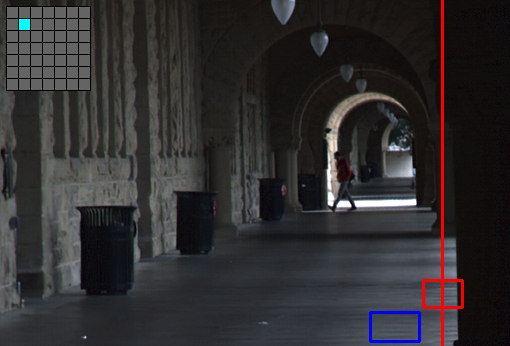}}
        & \multicolumn{2}{l}{\includegraphics[width=0.19\textwidth, height=0.130\textwidth]{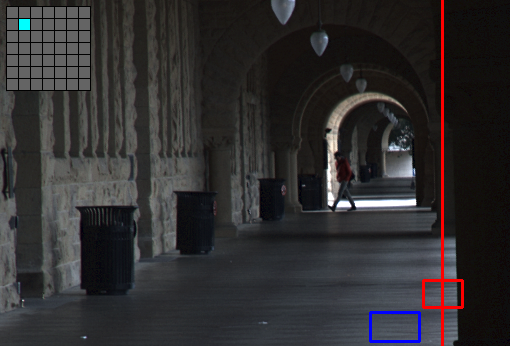}}
        &
        \\
        
        & \multicolumn{1}{l}{\includegraphics[width=0.09\textwidth, height=0.052\textwidth,cfbox=red 1pt 0pt] {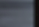}}
        & \multicolumn{1}{r}{\includegraphics[width=0.09\textwidth, height=0.052\textwidth,cfbox=blue 1pt 0pt]{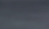}}
        & \multicolumn{1}{l}{\includegraphics[width=0.09\textwidth, height=0.052\textwidth,cfbox=red 1pt 0pt] {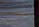}}
        & \multicolumn{1}{r}{\includegraphics[width=0.09\textwidth, height=0.052\textwidth,cfbox=blue 1pt 0pt]{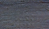}}
        & \multicolumn{1}{l}{\includegraphics[width=0.09\textwidth, height=0.052\textwidth,cfbox=red 1pt 0pt] {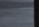}}
        & \multicolumn{1}{r}{\includegraphics[width=0.09\textwidth, height=0.052\textwidth,cfbox=blue 1pt 0pt]{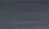}}
        & \multicolumn{1}{l}{\includegraphics[width=0.09\textwidth, height=0.052\textwidth,cfbox=red 1pt 0pt] {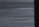}}
        & \multicolumn{1}{r}{\includegraphics[width=0.09\textwidth, height=0.052\textwidth,cfbox=blue 1pt 0pt]{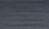}}
        & \multicolumn{1}{l}{\includegraphics[width=0.09\textwidth, height=0.052\textwidth,cfbox=red 1pt 0pt] {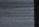}}
        & \multicolumn{1}{r}{\includegraphics[width=0.09\textwidth, height=0.052\textwidth,cfbox=blue 1pt 0pt]{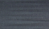}}
        &
        \\

        \multicolumn{1}{c}{}
        & \multicolumn{2}{l}{\includegraphics[width=0.19\textwidth, height=0.016\textheight]{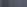}}
        & \multicolumn{2}{l}{\includegraphics[width=0.19\textwidth, height=0.016\textheight]{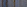}}
        & \multicolumn{2}{l}{\includegraphics[width=0.19\textwidth, height=0.016\textheight]{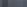}}
        & \multicolumn{2}{l}{\includegraphics[width=0.19\textwidth, height=0.016\textheight]{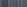}}
        & \multicolumn{2}{l}{\includegraphics[width=0.19\textwidth, height=0.016\textheight]{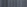}}
        &
        \\

        \multicolumn{1}{c}{}
        & \multicolumn{2}{l}{\includegraphics[width=0.19\textwidth, height=0.016\textheight]{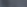}}
        & \multicolumn{2}{l}{\includegraphics[width=0.19\textwidth, height=0.016\textheight]{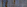}}
        & \multicolumn{2}{l}{\includegraphics[width=0.19\textwidth, height=0.016\textheight]{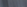}}
        & \multicolumn{2}{l}{\includegraphics[width=0.19\textwidth, height=0.016\textheight]{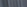}}
        & \multicolumn{2}{l}{\includegraphics[width=0.19\textwidth, height=0.016\textheight]{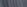}}
        &
        \\

        \multicolumn{1}{c}{\vspace{0.15cm}}
        & \multicolumn{2}{c}{33.08/0.9024}
        & \multicolumn{2}{c}{28.05/0.8669}
        & \multicolumn{2}{c}{42.59/0.9927}
        & \multicolumn{2}{c}{42.56/0.9926}
        & \multicolumn{2}{c}{}
        \\

    \end{tabular}
    }
    \caption{Visualization of DKNet and TE-DKNet. Three samples are presented in three rows respectively. In each row, a SAI, two zoomed-in views (red and blue boxes), two EPIs (along the red line and two angular dimensions in the red box), and PSNR/SSIM are given.}
    \label{fig:Qual2}
\end{figure*}

\subsubsection{Dense and Raw Image Connections}

As mentioned in Section \ref{section:Network Architecture}, the dense and raw image connections are employed in DKNet for information flow improvement. To verify their effects, we compared the DKNet variants with and without them. Apart from $\mathcal{K}^{\gamma}$, the same procedure is performed on $\mathcal{K}^{SAS}$ to factor out the effect of sophisticated decomposition kernels. We conduct the experiment in scale $4 \times$, and the results are depicted with their speed and trainable parameter numbers in Fig. \ref{fig:PerfNCost} (b) to demonstrate their efficiency. For simplicity, the dense-connection-only models are denoted as D, the raw-image-connection-only ones are denoted as R, and the full model is denoted as DR. The models with neither dense connections nor raw image connections are not included in the plots as they are malfunctioned producing out-of-range PSNR results.

Generally speaking, removing either of them leads to evident performance deterioration. The models without raw image connections are about 0.55 dB inferior to the full versions, from 36.07 dB to 35.52 dB in $\mathcal{K}^{\gamma}$ and from 35.55 dB to 35.01 dB in $\mathcal{K}^{SAS}$. The decrease is more serious if removing dense connections, causing 1.27 dB loss to $\mathcal{K}^{\gamma}$ from 36.07 dB to 34.80 dB, and 1.05 dB loss to $\mathcal{K}^{SAS}$ from 35.55 dB to 34.50 dB. The consequence of removing both types of connections is catastrophic, bringing down the performance to about 15.54 dB. This result has been anticipated as the models with 18 decomposition kernels are very deep, which are prone to failure due to gradient vanishment. As mentioned in Section \ref{section:Methodology}, this problem was previously tackled by using residual connections to support the skeleton \cite{yeungSAS_LFSR2019}, while we use the two different kinds of connections in DKNet.

The raw image connection itself turns up to be an economical alternative as it can support the deep network architecture but comes at much lower computation and memory costs. It is observed that both kernels with only raw image connections (R) run faster than the SAS version of Yeung \textit{et al.} \cite{yeungSAS_LFSR2019} while outperforming its 4D version in PSNR. Looking at the model with only dense connections (D), it can be found that removing raw image connections greatly impairs the performance but brings insignificant reduction of run-time and trainable parameters.

\subsubsection{Decomposition Kernels Numbers}
To discover the impact of stacking decomposition kernels, we train DKNet with different $L$ from 2 to 19, and the results are shown with their speed and trainable parameter numbers in Fig. \ref{fig:PerfNCost} (c). Continuous performance improvement is recorded from 34.59 dB to 36.07 dB when increasing $L$ from 2 to 18, suggesting the advancement of exploiting more decomposition kernels to obtain a larger receptive field in the sub-spaces and extract deeper spatio-angular features. However, the performance does not increase when $L > 18$, which means that the positive effect brought by stacking decomposition kernels has saturated, and over-fitting has been incurred.

\begin{figure*}[t]
    \centering
    \tabcolsep=0.03cm
    \renewcommand{\arraystretch}{0.6}
    \begin{tabu}{cccccccccc}
        \multicolumn{1}{l}{}
        &\multicolumn{1}{l}{}
        & \multicolumn{2}{c}{LR}
        & \multicolumn{2}{c}{DKNet}
        & \multicolumn{2}{c}{TE-DKNet}
        & \multicolumn{2}{c}{Ground-truth}
        \\
        
        \multirow{3}{*}{\rotatebox{90}{\hspace{33pt}{(a) Fabric}}}
        & \multirow{3}{*}[0.61cm]{\includegraphics[width=0.145\textwidth, height=0.100\textwidth]{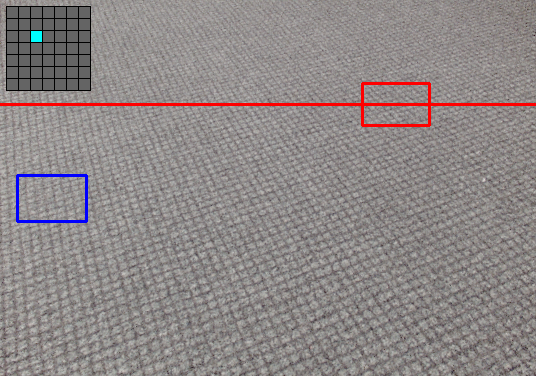}}
        & \multicolumn{1}{l}{\includegraphics[width=0.090\textwidth, height=0.052\textwidth,cfbox=red 1pt 0pt] {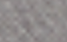}}
        & \multicolumn{1}{r}{\includegraphics[width=0.090\textwidth, height=0.052\textwidth,cfbox=blue 1pt 0pt]{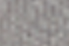}}
        & \multicolumn{1}{l}{\includegraphics[width=0.090\textwidth, height=0.052\textwidth,cfbox=red 1pt 0pt] {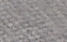}}
        & \multicolumn{1}{r}{\includegraphics[width=0.090\textwidth, height=0.052\textwidth,cfbox=blue 1pt 0pt]{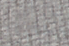}}
        & \multicolumn{1}{l}{\includegraphics[width=0.090\textwidth, height=0.052\textwidth,cfbox=red 1pt 0pt] {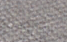}}
        & \multicolumn{1}{r}{\includegraphics[width=0.090\textwidth, height=0.052\textwidth,cfbox=blue 1pt 0pt]{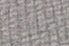}}
        & \multicolumn{1}{l}{\includegraphics[width=0.090\textwidth, height=0.052\textwidth,cfbox=red 1pt 0pt] {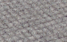}}
        & \multicolumn{1}{r}{\includegraphics[width=0.090\textwidth, height=0.052\textwidth,cfbox=blue 1pt 0pt]{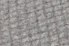}}
        \\

        &
        & \multicolumn{2}{l}{\includegraphics[width=0.192\textwidth, height=0.014\textheight]{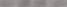}}
        & \multicolumn{2}{l}{\includegraphics[width=0.192\textwidth, height=0.014\textheight]{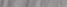}}
        & \multicolumn{2}{l}{\includegraphics[width=0.192\textwidth, height=0.014\textheight]{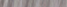}}
        & \multicolumn{2}{l}{\includegraphics[width=0.192\textwidth, height=0.014\textheight]{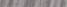}}
        \\

        &
        & \multicolumn{2}{l}{\includegraphics[width=0.192\textwidth, height=0.014\textheight]{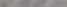}}
        & \multicolumn{2}{l}{\includegraphics[width=0.192\textwidth, height=0.014\textheight]{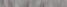}}
        & \multicolumn{2}{l}{\includegraphics[width=0.192\textwidth, height=0.014\textheight]{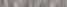}}
        & \multicolumn{2}{l}{\includegraphics[width=0.192\textwidth, height=0.014\textheight]{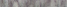}}
        \\

        \vspace{-0.25cm} \\

        \multirow{3}{*}{\rotatebox{90}{\hspace{32pt}{(b) Leather}}}
        & \multirow{3}{*}[0.61cm]{\includegraphics[width=0.145\textwidth, height=0.100\textwidth]{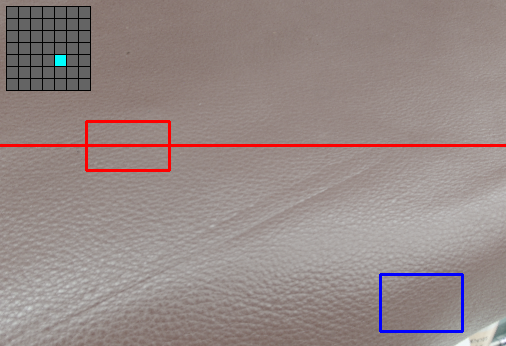}}
        & \multicolumn{1}{l}{\includegraphics[width=0.090\textwidth, height=0.052\textwidth,cfbox=red 1pt 0pt] {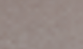}}
        & \multicolumn{1}{r}{\includegraphics[width=0.090\textwidth, height=0.052\textwidth,cfbox=blue 1pt 0pt]{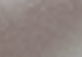}}
        & \multicolumn{1}{l}{\includegraphics[width=0.090\textwidth, height=0.052\textwidth,cfbox=red 1pt 0pt] {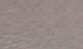}}
        & \multicolumn{1}{r}{\includegraphics[width=0.090\textwidth, height=0.052\textwidth,cfbox=blue 1pt 0pt]{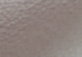}}
        & \multicolumn{1}{l}{\includegraphics[width=0.090\textwidth, height=0.052\textwidth,cfbox=red 1pt 0pt] {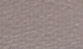}}
        & \multicolumn{1}{r}{\includegraphics[width=0.090\textwidth, height=0.052\textwidth,cfbox=blue 1pt 0pt]{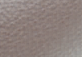}}
        & \multicolumn{1}{l}{\includegraphics[width=0.090\textwidth, height=0.052\textwidth,cfbox=red 1pt 0pt] {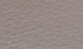}}
        & \multicolumn{1}{r}{\includegraphics[width=0.090\textwidth, height=0.052\textwidth,cfbox=blue 1pt 0pt]{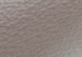}}
        \\

        &
        & \multicolumn{2}{l}{\includegraphics[width=0.192\textwidth, height=0.014\textheight]{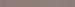}}
        & \multicolumn{2}{l}{\includegraphics[width=0.192\textwidth, height=0.014\textheight]{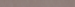}}
        & \multicolumn{2}{l}{\includegraphics[width=0.192\textwidth, height=0.014\textheight]{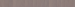}}
        & \multicolumn{2}{l}{\includegraphics[width=0.192\textwidth, height=0.014\textheight]{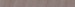}}
        \\

        &
        & \multicolumn{2}{l}{\includegraphics[width=0.192\textwidth, height=0.014\textheight]{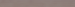}}
        & \multicolumn{2}{l}{\includegraphics[width=0.192\textwidth, height=0.014\textheight]{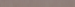}}
        & \multicolumn{2}{l}{\includegraphics[width=0.192\textwidth, height=0.014\textheight]{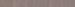}}
        & \multicolumn{2}{l}{\includegraphics[width=0.192\textwidth, height=0.014\textheight]{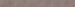}}
        \\
        
        \vspace{-0.25cm} \\

        \multirow{3}{*}{\rotatebox{90}{\hspace{32pt}{(c) Wood}}}
        & \multirow{3}{*}[0.61cm]{\includegraphics[width=0.145\textwidth, height=0.100\textwidth]{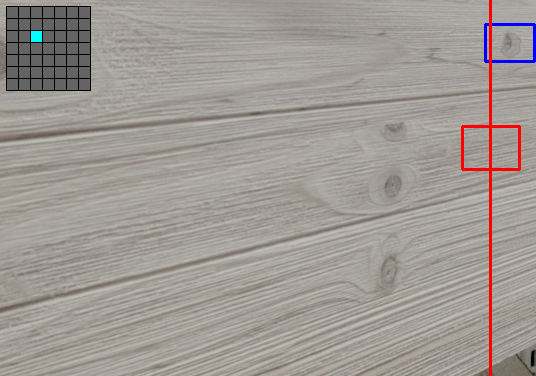}}
        & \multicolumn{1}{l}{\includegraphics[width=0.090\textwidth, height=0.052\textwidth,cfbox=red 1pt 0pt] {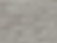}}
        & \multicolumn{1}{r}{\includegraphics[width=0.090\textwidth, height=0.052\textwidth,cfbox=blue 1pt 0pt]{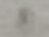}}
        & \multicolumn{1}{l}{\includegraphics[width=0.090\textwidth, height=0.052\textwidth,cfbox=red 1pt 0pt] {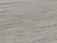}}
        & \multicolumn{1}{r}{\includegraphics[width=0.090\textwidth, height=0.052\textwidth,cfbox=blue 1pt 0pt]{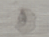}}
        & \multicolumn{1}{l}{\includegraphics[width=0.090\textwidth, height=0.052\textwidth,cfbox=red 1pt 0pt] {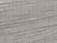}}
        & \multicolumn{1}{r}{\includegraphics[width=0.090\textwidth, height=0.052\textwidth,cfbox=blue 1pt 0pt]{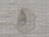}}
        & \multicolumn{1}{l}{\includegraphics[width=0.090\textwidth, height=0.052\textwidth,cfbox=red 1pt 0pt] {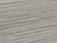}}
        & \multicolumn{1}{r}{\includegraphics[width=0.090\textwidth, height=0.052\textwidth,cfbox=blue 1pt 0pt]{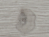}}
        \\

        &
        & \multicolumn{2}{l}{\includegraphics[width=0.192\textwidth, height=0.014\textheight]{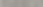}}
        & \multicolumn{2}{l}{\includegraphics[width=0.192\textwidth, height=0.014\textheight]{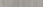}}
        & \multicolumn{2}{l}{\includegraphics[width=0.192\textwidth, height=0.014\textheight]{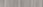}}
        & \multicolumn{2}{l}{\includegraphics[width=0.192\textwidth, height=0.014\textheight]{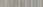}}
        \\

        &
        & \multicolumn{2}{l}{\includegraphics[width=0.192\textwidth, height=0.014\textheight]{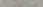}}
        & \multicolumn{2}{l}{\includegraphics[width=0.192\textwidth, height=0.014\textheight]{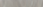}}
        & \multicolumn{2}{l}{\includegraphics[width=0.192\textwidth, height=0.014\textheight]{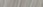}}
        & \multicolumn{2}{l}{\includegraphics[width=0.192\textwidth, height=0.014\textheight]{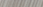}}
        \\

    \end{tabu}

    \caption{Visualization of LF material recognition. Two EPIs are extracted along the red line and the two angular dimensions in the red box separately.}
    \label{fig:QualRecg}
\end{figure*}

Regarding the costs, the trend is evident that reducing $L$ leads to a faster and lighter model at the cost of performance from 18 to 2. It is noticeable that the models with $L < 12$ run faster than the SAS version of Yeung \textit{et al.} \cite{yeungSAS_LFSR2019}, and the ones with $L < 5$ run even faster than the fastest competitor, resLF \cite{zhangResLF_CVPR2019}, while all of the variants outperform their competitors in terms of PSNR considerably. Significantly, the smallest model $L=2$ achieves the fastest speed, approximately 91 seconds per LF image, and has a similar model size with resLF \cite{zhangResLF_CVPR2019}, while it outperforms the heaviest method, the 4D version of Yeung \textit{et al.} \cite{yeungSAS_LFSR2019}, by 0.32 dB, indicating the superiority of DKNet's network design.

\subsubsection{Decomposition Types and Numbers}
As mentioned before, connection diversity plays an important role in DKNet. Since different types of kernels contain different numbers of convolutional layers, it is unfair to compare them given the same number $L$. In this section, I conduct an experiment to compare the models with the same number of convolutional layers. The result is shown in Fig. \ref{fig:PerfNCost} (d). 

Two baselines, $\mathcal{K}^{SAS} (L=18)$ and $(L=12)$ are set up. Corresponding to these two baselines, we test two four-connection kernels, $\mathcal{K}^{\alpha}$ and $\mathcal{K}^{\beta}$, and one six-connection kernel, $\mathcal{K}^{\gamma}$. To make the convolution number consistent, $L$ of $\mathcal{K}^{\alpha}$ and $\mathcal{K}^{\beta}$ is set to 9 and 6 for the two baselines respectively, and the one of $\mathcal{K}^{\gamma}$ is set to 6 and 4 likewise. Consequently, the group of $\mathcal{K}^{SAS} (L=18)$ (purple) has 36 convolutional layers within the kernels while the second group (orange) has 24 ones. 

In Fig. \ref{fig:PerfNCost} (d), it can be seen that, in each group, the three variants achieve very similar performance of PSNR. However, it should be noted that the four- and six-connection kernels are considerably efficient as they use only 1/4 to 1/3 of running time and their model sizes are between 60\% and 80\%. 

To further demonstrate the efficiency of decomposition kernels, we add one more variant to each group, $\mathcal{K}^{\gamma} (L=14)$ and $(L=9)$, which have similar parameter numbers with $\mathcal{K}^{SAS} (L=18)$ and $(L=12)$ respectively to discover how large the advantage can be brought by connection diversity given the same model sizes. As per the figure, a significant improvement is observed as the two models surpass their baseline by 0.35 dB and 0.61 dB, and their speed is around 30\% faster.

The result clearly verifies the effectiveness of decomposition kernels as having more diverse connections improves the model performance more than simply adding homogeneous convolutional layers.

\subsection{Study of Feature Loss} \label{section:StudyOfFeatureLoss}
To pursue visually pleasing results, we train DKNet with the proposed LFVGG loss in Eq. \ref{eq:final loss} and denote it as the Texture-Enhanced DKNet (TE-DKNet). Instead of training from scratch, we initialize the network with parameters trained by pixel-wise MSE in Eq. \ref{eq:pixel loss} to avoid the occurrence of an unstable training process. The output of layer \textit{relu5\_4} of VGG-19 is selected as the function $\phi_l$ and the weight $\lambda$ is set to $5e-4$. The experiments are conducted in the challenging $4 \times$ LFSR scale hereafter.

\subsubsection{Qualitative Results}
Although it is anticipated that $\mathcal{L}_{LFVGG}$ will lower TE-DKNet's PSNR and SSIM as happened in SISR \cite{johnsonVGGLoss_ECCV2016, ledigSRGAN_CVPR2017, SajjadiEnhanceNet_ICCV2017}, recording 35.91 dB PSNR, 0.16 dB lower than the original DKNet, some samples are found with better visual results in human vision as depicted in Fig. \ref{fig:Qual2}. SRGAN \cite{ledigSRGAN_CVPR2017} is also exhibited as the representative of perception-based SISR methods.

In \textit{general\_17}, the textures on the bricks are enhanced by TE-DKNet when compared to the original DKNet. Its EPIs are also sharper and closer to the ground-truth whereas the original DKNet produces blurry EPIs. Similar results can be observed in \textit{general\_45} where the bricks' textures of the road surface in the red and blue boxes are recovered richer and sharper. Although these textures do not correspond to the location in the ground-truth accurately, the super-resolved images are more visually pleasing. It is remarkable that TE-DKNet not only produces sharper patterns in both EPIs but also rectifies the pattern orientation in the second EPIs, which implies its correct LF structure to the ground-truth. Meanwhile, SISR-based SRGAN produces some details as well, however, since it cannot exploit the LF correlation information, not only its SAIs but also its EPIs are noisy and unnatural. What is worse, the relatively low darkness of \textit{general\_17} triggers SRGAN to render the image abnormally bright. 

One may question that these details are essentially learned noises as they appear to be somewhat random. However, we find that TE-DKNet could enhance not only irregular textures but also the regular ones, \textit{e.g.} the floor patterns in the red and blue boxes of \textit{general\_48} where TE-DKNet produces more strip-like textures with sharper edges while SRGAN produces noises. Likewise, the EPIs of TE-DKNet have more information and stronger conformity than the original DKNet, indicating the produced details are meaningful. Given these visual improvements, its PSNR and SSIM are very close to the original DKNet, suggesting that the visual enhancement of $\mathcal{L}_{LFVGG}$ does not necessarily impair TE-DKNet's performance under the conventional pixel-wise metrics.

\subsubsection{Quantitative Results by Material Recognition}
\begin{table}[t!]
    \caption{Results of light field material recognition.}
    \label{tab:PerfRecg}
    \centering
    \begin{tabular}{|l|c|}
        \hline
        Model                       & Accuracy  \\\hline
        LR                          & 71.39\%   \\\hline
        DKNet                       & 75.83\%   \\\hline
        TE-DKNet                    & 77.50\%   \\\hline
        Ground-truth                & 77.78\%   \\\hline
    \end{tabular}
\end{table}
To better assess the visual quality of reconstructed images and tackle the inherent deficiencies of pixel-based metrics, some alternative evaluation methods are developed to provide quantitative measures that correlate better with human perception, \textit{e.g.} Sajjadi \textit{et al.} \cite{SajjadiEnhanceNet_ICCV2017} and Park \textit{et al.} \cite{parkSRFeat_ECCV2018} proposed an indirect metric using object recognition for this purpose. Inspired by them, we propose to use the recognition model to assess the perceptual quality of LF images. Instead of directly adopting single-image recognition models, in order to take both human perception and LF structure into consideration, we resort to LF material recognition by following Wang \textit{et al.} \cite{wangLFRecognition_ECCV2016} to train a classification model for this task. They proposed a dataset containing 12 categories of materials including fabric, leather, and wood. Each category has 100 LF images featuring an abundance of material textures. Following the evaluation protocol of \cite{wangLFRecognition_ECCV2016}, we split each category into 70\% and 30\% of the samples for training and testing correspondingly.

The results are reported in Table \ref{tab:PerfRecg}. The trained model has achieved 77.78\% accuracy on the original test set (ground-truth). As the material textures are seriously damaged in the down-sampled LR images, the accuracy decreases by 6.39\% to 71.39\%. With the aid of DKNet, the accuracy recovers partially by 4.44\% to 75.83\%. TE-DKNet improves the accuracy further by 1.67\% to 77.50\%, which is very close to the ground-truth. Evidently, the 1.67\% increase is contributed by the LFVGG loss as it enhances the textures. In light of this, it would be a convenient approach to exploit either DKNet or TE-DKNet for pre-processing to improve the accuracy of LF material recognition.

Some representative samples are depicted in Fig. \ref{fig:QualRecg}. The visual results demonstrate the significant enhancement of textures made by TE-DKNet. In Fig. \ref{fig:QualRecg} (a) and (b), the textures of fabric and leather are particularly dense or irregular. It is observed that the original DKNet only recovers them partially while TE-DKNet produces more textures that do not perfectly match the ground-truth pixel-wise but look realistic. In Fig. \ref{fig:QualRecg} (c), the textures of wood are more intact and sharper using TE-DKNet. It is remarkable when looking at the second EPIs where the original DKNet generates blurry and wrongly oriented patterns, but with the aid of LFVGG loss, the patterns are sharpened and the orientation is rectified. This phenomenon also conforms to the observation in Fig. \ref{fig:QualKernels} that EPIs of both angular dimensions can provide insights into the reconstructed LF images from another perspective.

It is worth noticing that TE-DKNet is not trained on the material samples but the original 130 LF images of general scenes used in the previous experiments. Under such a circumstance, DKNet is still competent to produce rich textures and boost the accuracy of LF material recognition, manifesting DKNet's generalization ability across different datasets.

\begin{table*}[ht]
    \centering

    \caption{The network architecture of DKNet.}

    \resizebox{\textwidth}{!}{
    
    \begin{tabular}{|l|c|c|c|c|c|c|}
    \hline
    & Kernel Size / Remark  &  Input Size & Output Size & Stride & Pad & \#Parameters \\
    \hline\hline
    
    \bf LR $X$ & - & - & $(8,8,3,32,32)$ & - & - & - \\
    \hline\hline
    
    \multicolumn{7}{|l|}{\bf Feature Extraction} \\
    \hline
    Initial Convolution      & - & - & - & - & - & - \\
    - Reshape to $(x, y)$         & -                       & $(8,8,3,32,32)$   &  $(64,3,32,32)$       & -     & -     & -                 \\
    - Convolution on $(x, y)$     & $(3, 32, 3, 3)$         & $(64,3,32,32)$    & $(64,32,32,32)$       & 1/1   & 1/1   & 896               \\
    - Reshape to Normal           & -                       & $(64,32,32,32)$   & $(8,8,32,32,32)$      & -     & -     & -                 \\\hline
    
    Decomposition Kernel $\mathcal{K}^{\gamma}_1$           & - & - & - & - & - & *56352 \\
    - Concatenation               & Features                & $(8,8,32,32,32)$  & $(8,8,35,32,32)$      & -     & -     & -                 \\
                                  & Raw Image Connections   & $(8,8,3,32,32)$   &                       & -     & -     & -                 \\
    - Reshape to $(x, y)$         & -                       & $(8,8,35,32,32)$  & $(64,35,32,32)$       & -     & -     & -                 \\
    - Convolution on $(x, y)$     & $(35, 32, 3, 3)$        & $(64,35,32,32)$   & $(64,32,32,32)$       & 1/1   & 1/1   & 10,112            \\
    - Reshape to $(u, v)$         & -                       & $(64,32,32,32)$   & $(1024,3,8,8)$        & -     & -     & -                 \\
    - Convolution on $(u, v)$     & $(32, 32, 3, 3)$        & $(1024,32,8,8)$   & $(1024,32,8,8)$       & 1/1   & 1/1   & 9,248             \\
    - Reshape to $(u, x)$         & -                       & $(1024,32,8,8)$   & $(256,32,8,32)$       & -     & -     & -                 \\
    - Convolution on $(u, x)$     & $(32, 32, 3, 3)$        & $(256,32,8,32)$   & $(256,32,8,32)$       & 1/1   & 1/1   & 9,248             \\
    - Reshape to $(v, y)$         & -                       & $(256,32,8,32)$   & $(256,32,8,32)$       & -     & -     & -                 \\
    - Convolution on $(v, y)$     & $(32, 32, 3, 3)$        & $(256,32,8,32)$   & $(256,32,8,32)$       & 1/1   & 1/1   & 9,248             \\
    - Reshape to $(u, y)$         & -                       & $(256,32,8,32)$   & $(256,32,8,32)$       & -     & -     & -                 \\
    - Convolution on $(u, y)$     & $(32, 32, 3, 3)$        & $(256,32,8,32)$   & $(256,32,8,32)$       & 1/1   & 1/1   & 9,248             \\
    - Reshape to $(v, x)$         & -                       & $(256,32,8,32)$   & $(256,32,8,32)$       & -     & -     & -                 \\
    - Convolution on $(v, x)$     & $(32, 32, 3, 3)$        & $(256,32,8,32)$   & $(256,32,8,32)$       & 1/1   & 1/1   & 9,248             \\
    - Reshape to Normal           & -                       & $(256,32,8,32)$   & $(8,8,32,32,32)$      & -     & -     & -                 \\\hline
    
    Decomposition Kernel $\mathcal{K}^{\gamma}_2$           & - & - & - & - & - & *56352 \\
    - Concatenation               & Features                & $(8,8,32,32,32)$  & $(8,8,35,32,32)$      & -     & -     & -                 \\
                                  & Raw Image Connections   & $(8,8,3,32,32)$   &                       & -     & -     & -                 \\
    - Reshape to $(x, y)$         & -                       & $(8,8,35,32,32)$  & $(64,35,32,32)$       & -     & -     & -                 \\
    - Convolution on $(x, y)$     & $(35, 32, 3, 3)$        & $(64,35,32,32)$   & $(64,32,32,32)$       & 1/1   & 1/1   & 10,112            \\
    - Reshape \& Convolution $\times 5$     & $\cdots$     & $\cdots$          & $\cdots$        & $\cdots$ & $\cdots$ & 9,248$\times 5$   \\
    - Reshape to Normal           & -                       & $(256,32,8,32)$   & $(8,8,32,32,32)$      & -     & -     & -                 \\\hline
    
    Decomposition Kernel $\mathcal{K}^{\gamma}_3$           & - & - & - & - & - & *65568 \\
    - Concatenation               & Features                & $(8,8,32,32,32)$  & $(8,8,67,32,32)$      & -     & -     & -                 \\
                                  & Raw Image Connections   & $(8,8,3,32,32)$   &                       & -     & -     & -                 \\
                                  & Dense Connections       & $(8,8,32,32,32)$  &                       & -     & -     & -                 \\
    - Reshape to $(x, y)$         & -                       & $(8,8,67,32,32)$  & $(64,67,32,32)$       & -     & -     & -                 \\
    - Convolution on $(x, y)$     & $(67, 32, 3, 3)$        & $(64,67,32,32)$   & $(64,32,32,32)$       & 1/1   & 1/1   & 19,328            \\
    - Reshape \& Convolution $\times 5$     & $\cdots$     & $\cdots$          & $\cdots$        & $\cdots$ & $\cdots$ & 9,248$\times 5$   \\
    - Reshape to Normal           & -                       & $(256,32,8,32)$   & $(8,8,32,32,32)$      & -     & -     & -                 \\\hline
    
    Decomposition Kernel $\mathcal{K}^{\gamma}_4$           & - & - & - & - & - & *74784 \\
    - Concatenation               & Features                & $(8,8,32,32,32)$  & $(8,8,99,32,32)$      & -     & -     & -                 \\
                                  & Raw Image Connections   & $(8,8,3,32,32)$   &                       & -     & -     & -                 \\
                                  & Dense Connections       & $(8,8,64,32,32)$  &                       & -     & -     & -                 \\
    - Reshape to $(x, y)$         & -                       & $(8,8,99,32,32)$  & $(64,99,32,32)$       & -     & -     & -                 \\
    - Convolution on $(x, y)$     & $(99, 32, 3, 3)$        & $(64,99,32,32)$   & $(64,32,32,32)$       & 1/1   & 1/1   & 28,544            \\
    - Reshape \& Convolution $\times 5$     & $\cdots$     & $\cdots$          & $\cdots$        & $\cdots$ & $\cdots$ & 9,248$\times 5$   \\
    - Reshape to Normal           & -                       & $(256,32,8,32)$   & $(8,8,32,32,32)$      & -     & -     & -                 \\\hline
    
    Decomposition Kernel $\mathcal{K}^{\gamma}_5$           & - & - & - & - & - & *84000  \\\hline
    $\cdots$                                                & $\cdots$ & $\cdots$ & $\cdots$ & $\cdots$ & $\cdots$ & $\cdots$ \\\hline
    Decomposition Kernel $\mathcal{K}^{\gamma}_{18}$        & - & - & - & - & - & *203808  \\\hline\hline

    \multicolumn{7}{|l|}{\bf Image Generation} \\
    \hline
    Concatenation                                             & Features                  & $(8,8,32\times18,32,32)$    & $(8,8,579,32,32)$     & -     & -     & -             \\
                                                              & Raw Image Connections     & $(8,8,3,32,32)$             &                       & -     & -     & -             \\\hline
    Feature Reduction                                         & - & - & - & - & - &  \\
    - Reshape to $(u, v)$                                     & -                         & $(8,8,579,32,32)$           & $(1024,579,8,8)$      & -     & -     & -             \\
    - Convolution on $(u, v)$                                 & $(579, 32, 3, 3)$         & $(1024,579,8,8)$            & $(1024,32,4,4)$       & 2/2   & 1/1   & 166,784       \\
    Feature Reduction (Convolution on $(u, v)$)               & $(32, 3072, 4, 4)$        & $(1024,32,4,4)$             & $(1024,3072,1,1)$     & 1/1   & 0/0   & 1,575,936     \\
    Reshape                                                   & -                         & $(1024,3072,1,1)$           & $(8,8,48,32,32)$      & -     & -     & -             \\
    Pixel Shuffler                                            & -                         & $(8,8,48,32,32)$            & $(8,8,48,128,128)$    & -     & -     & -             \\\hline
    \end{tabular}
    }
    \label{tab:Archi}

    \begin{itemize}
        \item As an example, a network with 18 $\mathcal{K}^{\gamma}$ kernels ($L=18$) performs $4\times$ super-resolution on a LF image of size $(8,8,3,32,32)$.
        \item The kernel size is presented as (input channels, output channels, width, height).
        \item The input and output sizes are presented as $(U, V, C, W, H)$. At times, two of $U$, $V$, $W$ and $H$ are merged at to reduce the dimensions and perform convolutions.
        \item In the column of \#Parameters, * means the combination of sub-layers' parameters.
    \end{itemize}
\end{table*}

\section{Conclusion} \label{section:Conclusion}
In this paper, we have examined the decomposition operations of 4D LF sub-spaces, systematically unified them into the novel concept of decomposition kernels, and designed a series of new kernels with intra- and inter-domain connections to overcome the drawbacks existing in the other methods. We embedded these decomposition kernels in our proposed DKNet for LFSR. With dense connections and raw image connections, DKNet was able to extract deep and comprehensive spatio-angular features. The objective experimental results showed that DKNet has achieved state-of-the-art performance under the conventional pixel-wise PSNR and SSIM metrics. The ablation studies have discovered the characteristics of the proposed decomposition kernels and demonstrated the effectiveness of sufficient sub-space exploitation.

To further pursue visually pleasing results, particularly for realistic textures, we have proposed the novel LFVGG loss to enhance DKNet to tackle the drawback of pixel-wise losses. In addition, an indirect perception-based metric has been proposed to quantitatively evaluate the perceptual enhancement by smartly utilizing the LF material recognition model that correlates with both human perception and LF structure. Extensive experiments have been conducted to evaluate TE-DKNet in both objective and subjective studies. The subjective study has shown that TE-DKNet trained by the LFVGG loss obtained superior visual improvements over traditional methods, especially in the areas with rich textures. In the objective study, we also observed that TE-DKNet can boost the accuracy of LF material recognition, which confirms its outstanding capability in generating authentic textures.


\ifCLASSOPTIONcaptionsoff
\newpage
\fi

\bibliographystyle{IEEEtran}
\bibliography{bare_jrnl}

\end{document}